\documentclass{IEEEoj}
\usepackage{cite}
\usepackage{amsmath,amssymb,amsfonts}
\usepackage{algorithmic}
\usepackage{graphicx,color}
\usepackage{textcomp}
\def\BibTeX{{\rm B\kern-.05em{\sc i\kern-.025em b}\kern-.08em
    T\kern-.1667em\lower.7ex\hbox{E}\kern-.125emX}}

\usepackage[
    hidelinks,
]{hyperref}

\usepackage{graphicx}
\usepackage{tabularx}

\usepackage[
	ruled,
	noend,  
	linesnumbered,
]
{algorithm2e}
\SetKwComment{Comment}{}{}  

\usepackage[
    table,
]{xcolor}

\usepackage{xspace}

\usepackage{cite}

\usepackage{orcidlink}

\usepackage{relsize}

\usepackage{			
	amsmath,
	amssymb,
	amsfonts,
}

\usepackage{mathtools}

\usepackage{siunitx}

\usepackage{tikz}
\usetikzlibrary{
	positioning,				
	shapes.geometric,			
	decorations.pathreplacing,	
	shadows,					
	patterns,					
	math,
	arrows.meta,
	shapes.multipart,			
	external,
}

\usepackage[
	toc=False,
	numberline=False,
	style=tree,
	sort=standard,
	nopostdot=True,
	acronym,
    shortcuts=True,
]{glossaries}

\glsdisablehyper  

\newacronym{sota}{SotA}{State of the Art}
\newacronym
    [longplural={Degrees of Freedom}]
    {dof}{DoF}{Degree of Freedom}
    
\newacronym{qos}{QoS}{Quality of Service}

\newacronym{3gpp}{3GPP}{3rd Generation Partnership Program}

\newacronym{embb}{eMBB}{Enhanced Mobile Broadband}
\newacronym{urllc}{URLLC}{Ultra Reliable and Low Latency Communications}
\newacronym{mmtc}{mMTC}{Massive Machine Type Communications}

\newacronym{tx}{Tx}{Transmit}
\newacronym{rx}{Rx}{Receive}

\newacronym
    [longplural={Channel State Information at Transmitter}]
    {csit}{CSIT}{Channel State Information at Transmitter}
\newacronym
    [longplural={Channel State Information}]
    {csi}{CSI}{Channel State Information}
\newacronym{ber}{BER}{Bit Error Rate}
\newacronym{snr}{SNR}{Signal to Noise Ratio}
\newacronym{sinr}{SINR}{Signal to Interference and Noise Ratio}
\newacronym{slnr}{SLNR}{Signal to Leakage and Noise Ratio}

\newacronym{sgd}{SGD}{Stochastic Gradient Descent}
\newacronym{mmse}{MMSE}{Minimum Mean Squared Error}
\newacronym{zf}{ZF}{Zero Forcing}

\newacronym{dtft}{DTFT}{Discrete Time Fourier Transform}
\newacronym{dft}{DFT}{Discrete Fourier Transform}
\newacronym{fft}{FFT}{Fast Fourier Transform}

\newacronym{ofdm}{OFDM}{Orthogonal Frequency Division Multiplex}
\newacronym{ofdma}{OFDMA}{Orthogonal Frequency Division Multiple Access}
\newacronym{noma}{NOMA}{Non Orthogonal Multiple Access}
\newacronym{sdma}{SDMA}{Spatial Division Multiple Access}

\newacronym{siso}{SISO}{Single Input Single Output}
\newacronym{simo}{SIMO}{Single Input Multiple Output}
\newacronym{miso}{MISO}{Multiple Input Single Output}
\newacronym{mimo}{MIMO}{Multiple Input Multiple Output}

\newacronym{iui}{IUI}{Inter-User Interference}

\newacronym{mdp}{MDP}{Markov Decision Process}
\newacronym{ml}{ML}{Machine Learning}
\newacronym{dl}{DL}{Deep Learning}
\newacronym{rl}{RL}{Reinforcement Learning}

\newacronym{lr}{LR}{Learning Rate}

\newacronym{nn}{NN}{Neural Network}
\newacronym{cnn}{CNN}{Convolutional Neural Networks}
\newacronym{acnn}{AcNN}{Actor Neural Network}
\newacronym{crnn}{CrNN}{Critic Neural Network}

\newacronym{dqn}{DQN}{Deep Q Network}
\newacronym{ddpg}{DDPG}{Deep Deterministic Policy Gradient}
\newacronym{trpo}{TRPO}{Trust Region Policy Optimization}
\newacronym{ppo}{PPO}{Proximal Policy Optimization}
\newacronym{sac}{SAC}{Soft Actor-Critic}

\newacronym{ntn}{NTN}{Non-Terrestrial Networks}
\newacronym{leo}{LEO}{Low Earth Orbit}
\newacronym{geo}{GEO}{Geostationary Earth Orbit}

\newacronym
	[longplural={Angles of Departure}]
	{aod}{AoD}{Angle of Departure}
\newacronym{los}{LoS}{Line of Sight}
\newacronym{ula}{ULA}{Uniform Linear Array}

\newglossaryentry{ex}{%
	name={example},
	description={an example},
}%

\input{99_custommacros.sty}
\newcommand{\idx}						{i}
\newcommand{\numantennaspersat}			{N}
\newcommand{\antennaid}					{n}
\newcommand{\userid}					{k}
\newcommand{\numusers}					{K}
\newcommand{\satelliteindex}			{m}
\newcommand{\numsats}					{M}
\newcommand{\gaininternal}				{G}
\newcommand{\txdata}					{x}
\newcommand{\txsymbol}                  {u}
\newcommand{\rxdata}					{y}
\newcommand{\csiinternal}				{h}
\newcommand{\csimatrixinternal}			{H}
\newcommand{\precodinginternal}			{w}
\newcommand{\precodingmatrixinternal}	{W}
\newcommand{\noise}						{n}
\newcommand{\wavelength}				{\lambda}
\newcommand{\distance}					{d}
\newcommand{\phaseshift}				{\varphi}
\newcommand{\internalsteering}			{v}
\newcommand{\aod}						{\nu}
\newcommand{\largescalefading}			{\varsigma}
\newcommand{\internalstatevec}			{s}
\newcommand{\internalactionvec}			{a}
\newcommand{\expbuffersize}				{Z}
\newcommand{\expbufferindex}			{z}
\newcommand{\stepsperupdate}			{T}
\newcommand{\loss}						{\text{L}}
\newcommand{\error}						{\varepsilon}
\newcommand{\errorterm}					{\epsilon}
\newcommand{\txpower}					{P}
\newcommand{\fiddleparam}				{\zeta}
\newcommand{\steeringautocorrelation}	{\mathbf{R}}
\newcommand{\genericfunction}{f}

\newcommand{\usergain}{\gaininternal_{\text{Usr}}}
\newcommand{\satgain}{\gaininternal_{\text{Sat}}}
\newcommand{\csivector}{\mathbf{\csiinternal}}
\newcommand{\csimatrix}{\mathbf{\csimatrixinternal}}
\newcommand{\csivectorestim}{\mathbf{\tilde{\csiinternal}}}
\newcommand{\csimatrixestimsat}{\mathbf{\tilde{\tilde{\csimatrixinternal}}}}
\newcommand{\csimatrixestim}{\mathbf{\tilde{\csimatrixinternal}}}
\newcommand{\csimatrixestimdecone}{\mathbf{\tilde{\csimatrixinternal}}_{\text{L1}, \satelliteindex}}
\newcommand{\csimatrixestimdectwo}{\mathbf{\tilde{\csimatrixinternal}}_{\text{L2}, \satelliteindex}}
\newcommand{\precodingvector}{\mathbf{\precodinginternal}}
\newcommand{\precodingmatrix}{\mathbf{\precodingmatrixinternal}}
\newcommand{\userset}{\mathcal{\numusers}}
\newcommand{\satelliteset}{\mathcal{\numsats}}
\newcommand{\steeringvec}{\mathbf{\internalsteering}}
\newcommand{\steeringsca}{\internalsteering}
\newcommand{\largescalesigma}{\sigma_{\largescalefading}}
\newcommand{\statevec}{\mathbf{\internalstatevec}}
\newcommand{\actorfunction}{\boldsymbol{\internalactor}}
\newcommand{\paramsactor}{\boldsymbol{\internalparameter}_{\internalactor}}
\newcommand{\actoroutput}{\mathbf{\internalactionvec}'}
\newcommand{\actoroutputmeans}{\mathbf{\internalactionvec}_{\text{m}}'}
\newcommand{\actoroutputlogstd}{\mathbf{\internalactionvec}_{\sigma}'}
\newcommand{\actoroutputsampled}{\mathbf{\internalactionvec}}
\newcommand{\criticfunction}{\internalcritic}
\newcommand{\paramscritic}{\boldsymbol{\internalparameter}_{\internalcritic}}
\newcommand{\paramscriticnum}[1]{\boldsymbol{\internalparameter}_{\internalcritic, #1}}
\newcommand{\actoroutputsampledreal}{\mathbf{\internalactionvec}_{\text{Re}}}
\newcommand{\actoroutputsampledimag}{\mathbf{\internalactionvec}_{\text{Im}}}
\newcommand{\actoroutputsampledrealsca}{\internalactionvec_{\text{Re}, \idx}}
\newcommand{\actoroutputsampledimagsca}{\internalactionvec_{\text{Im}, \idx}}
\newcommand{\sumrate}{\internalreward}
\newcommand{\ergodicsumrate}{\bar{\internalreward}}
\newcommand{\wiggleusr}{\tilde{\distance}_{\userset}}
\newcommand{\wigglesat}{\tilde{\distance}_{\satelliteset}}
\newcommand{\batchset}{\mathcal{\batchsize}}
\newcommand{\erroraod}{\error_{\text{aod}, \userid, \satelliteindex}}
\newcommand{\erroraodbound}{\Delta\error_{\text{aod}}}
\newcommand{\errorphase}{\error_{\text{ph}, \userid, \satelliteindex}}
\newcommand{\errorphasevariance}{\variance_{\text{ph}}^{\num{2}}}
\newcommand{\errortermaod}{\boldsymbol{\errorterm}_{\text{aod}, \userid, \satelliteindex}}
\newcommand{\errortermphase}{\errorterm_{\text{ph}, \userid, \satelliteindex}}
\newcommand{\precodingmatrixmmse}{\precodingmatrix_{\text{MMSE}}}
\newcommand{\precodingvectorslnr}{\precodingvector_{\text{SLNR}, \userid}}
\newcommand{\precodingmatrixslnr}{\precodingmatrix_{\text{SLNR}}}
\newcommand{\learningrateactor}{\fiddleparam_{\internalactor}}
\newcommand{\learningratecritic}{\fiddleparam_{\internalcritic}}
\newcommand{\entropyscale}{\fiddleparam_{\text{e}}}
\newcommand{\weightregulscaleactor}{\fiddleparam_{\internalparameter_{\internalactor}}}
\newcommand{\weightregulscalecritic}{\fiddleparam_{\internalparameter_{\internalcritic}}}
\newcommand{\minsamples}{\expbuffersize'_{\text{min}}}
\newcommand{\maxeigslnr}{\boldsymbol{\fiddleparam}_{\userid, \text{max}}}
\newcommand{\powerchannelslnr}{\variance}
\AtBeginDocument{\definecolor{ojcolor}{cmyk}{0.93,0.59,0.15,0.02}}
\def\OJlogo{\vspace{-4pt}\hskip-4pt\includegraphics[height=18pt]{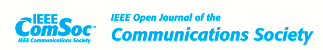}}
\begin{document}
\receiveddate{XX Month, XXXX}
\reviseddate{XX Month, XXXX}
\accepteddate{XX Month, XXXX}
\publisheddate{XX Month, XXXX}
\currentdate{11 January, 2024}
\doiinfo{OJCOMS.2024.011100}

\title{Model-Free Robust Beamforming in Satellite Downlink using Reinforcement Learning}

\author{Alea Schröder\IEEEauthorrefmark{1,2} \IEEEmembership{(Student Member, IEEE)}, Steffen Gracla \IEEEauthorrefmark{1,2}, Carsten Bockelmann \IEEEauthorrefmark{1} \IEEEmembership{(Member, IEEE}), Dirk Wübben \IEEEauthorrefmark{1,2}
\IEEEmembership{(Senior Member, IEEE)}, Armin Dekorsy \IEEEauthorrefmark{1,2} \IEEEmembership{(Senior Member, IEEE})}
\affil{Department of Communications Engineering, University of Bremen, 28359 Bremen, Germany}
\affil{Gauss-Olbers Space Technology Transfer Center, University of Bremen, 28359 Bremen, Germany}
\corresp{CORRESPONDING AUTHOR: Alea Schröder (e-mail: schroeder@ant.uni-bremen.de).}
\authornote{This work was partly funded by the German Federal Ministry of Research, Technology and Space (BMFTR) under grant 16KISK016 (Open6GHub) and the European Space Agency (ESA) under contract number 4000139559/22/UK/AL (AIComS)}
\markboth{Preparation of Papers for IEEE OPEN JOURNALS}{Author \textit{et al.}}


\begin{abstract}
	Satellite-based communications are expected to be a substantial future market in 6G networks. As satellite constellations grow denser and transmission resources remain limited, frequency reuse plays an increasingly important role in managing inter-user interference.
	In the multi-user downlink, precoding enables the reuse of frequencies across spatially separated users, greatly improving spectral efficiency.
	The analytical calculation of suitable precodings for perfect channel information is well studied, however, their performance can quickly deteriorate when faced with, e.g., outdated channel state information or, as is particularly relevant for satellite channels, when position estimates are erroneous.
    Deriving robust precoders under imperfect channel state information is not only analytically intractable in general but often requires substantial relaxations of the optimization problem or heuristic constraints to obtain feasible solutions.
	Instead, in this paper we flexibly derive robust precoding algorithms from given data using reinforcement learning.
	We describe how we adapt the applied Soft Actor-Critic learning algorithm to the problem of downlink satellite beamforming and show numerically that the resulting precoding algorithm adjusts to all investigated scenarios.
	The considered scenarios cover both single satellite and cooperative multi-satellite beamforming, using either global or local channel state information, and two error models that represent increasing levels of uncertainty.
	We show that the learned algorithms match or markedly outperform two analytical baselines in sum rate performance, adapting to the required level of robustness.
	We also analyze the mechanisms that the learned algorithms leverage to achieve robustness.
	The implementation is publicly available for use and reproduction of the results.
\end{abstract}


\glsresetall  
\maketitle%
%


\section{INTRODUCTION}
\label{sec:introduction}
\IEEEPARstart{T}{his} paper examines the use of deep \ac{rl} for the problem of beamforming in satellite-to-earth downlink.
In future mobile communications, satellites are expected to work in cooperation with traditional terrestrial networks, rather than being a competing mode of connection.
Satellite-empowered \ac{ntn} can bring benefits that complement the weaker points of terrestrial networks, primarily by offering direct connectivity to areas with low coverage, e.g., remote livings, ships at sea or aircrafts during flight, and disaster-stricken areas.
The development of widely accessible satellite communications was spurred in recent years by the advancements in the field of smaller, less expensive \ac{leo} satellites, with their lower orbit height between  \SI{600}{\km} and \SI{1000}{\km} offering short round-trip latency.
The satellite communications market is expected to grow rapidly, e.g., \cite{stanley2024satellite}~estimating the space industry to surge to a volume of \SI{1}[\$]{} Trillion by \num{2040}, over half of that from mobile connectivity applications.

In order to achieve these goals, spectral efficiency is an ongoing topic of discussion.
With competition for bandwidth and existing terrestrial network interference, satellites may opt to use beamforming to enable \ac{sdma}.
In beamforming, antenna arrays may be used to shape radiation patterns towards a desired spatial direction while suppressing interference in other directions.
This enables satellites to reuse the same frequency for multiple users in different positions, thereby increasing spectral efficiency.
Particularly with direct-to-handheld communications in mind, joint beamforming among multiple satellites allows very narrow transmission beams, separating users that are fairly close to each other.
Among others, we have shown in prior work~\cite{gracla_2023_learning} that beamforming \ac{sdma} can indeed lead to increased spectral efficiency in the \ac{ntn} context.

Mathematically, beamforming is performed by purposefully overlapping multiple simultaneous precoded transmissions, i.e., from multiple antennas, based on the available \ac{csit}~\cite{tse2005fundamentals}.
We describe this process in detail in the following section.
The calculation of optimal precodings is well known from terrestrial communications, and typical precoders such as \ac{mmse} precoding do a serviceable job when applied to \ac{ntn} conditions~\cite{chatzinotas2011energy}.
In satellite downlink, conventional \ac{csit} estimation based on uplink \ac{csi} can be inaccurate, as the uplink conditions may not reliably reflect the downlink channel. Therefore, we assume that the \ac{csit} in satellite downlink \ac{los} channels is primarily obtained from positioning information instead. However, such position-based \ac{csit} can become outdated quickly. This is due to the large distances, high satellite velocities, and synchronization delays. As a result, maintaining accurate channel knowledge becomes particularly challenging.
Using such erroneous or outdated channel information will quickly degrade the performance of typical precoding algorithms.
Conventionally, channel models can be extended with an error model that tracks reality as closely as possible, and new precoding algorithms tailored to these erroneous channels can be derived analytically, e.g., \cite{roper2023robust, bazzi2016robust, lin2020robust}.
However, these error models are merely an approximation of reality, and mathematical intractability may force a relaxation of the problem regardless.
This work looks particularly at the problem of sum rate maximization in the presence of erroneous \ac{csit}, which is known to be NP-hard~\cite{roper2023robust}.
Further, exceeding computational complexity may preclude a precoder from consideration in the context of satellite communications even if it achieves better sum rate performance than less complex algorithms~\cite{perez2019signal}. 

More recently, data-driven \ac{ml} has become an attractive option to derive algorithms from observations rather than from a model.
\ac{ml} iteratively tries to find a best fit to given data, whatever the underlying real processes and error sources may be.
This approach has been shown, both within the communications domain~\cite{zhang2019deep, naous2023reinforcement} as well as other domains, to match or outperform traditional model-based algorithms on these types of problems.
In the context of satellite precoding, e.g., \cite{liu2022robust} have used \ac{ml} to approximate a lower complexity implementation of an existing robust precoder using supervised learning.

In this paper, we use the subdomain of \ac{rl}~\cite{sutton2018reinforcement} to tackle the problem of downlink satellite precoding.
\ac{rl}, as described in depth in subsequent sections, starts with a random precoding strategy and iteratively improves upon it by a process of scientific trial-and-error.
Specifically, a learning agent will select a precoding for a given \ac{csit} estimate, and receive a resulting performance metric, i.e., the sum rate.
Based on such repeated interactions, the learning agent adapts its precoding strategy to promote precodings that lead to high sum rates and demote decisions that lead to low sum rates.
This carries the major advantage that, in contrast to supervised learning, no pre-existing precoding algorithms, system models or pre-labeled training data are required to learn.
Conversely, the problem of selecting precodings is also a particularly good fit for  \ac{rl} as there are no long term dependencies on the decisions, i.e., selecting a precoding at one time~\( \timeindex \) will not influence future channel states at times~\(\timeindex+\timeindex'\).
This is beneficial since long term time dependencies are a major factor of instability for \ac{rl} convergence.
In \cite{alsenwi2023robust}, the authors have applied a similar \ac{rl} approach on a mixed \ac{qos} optimization objective using codebook style precoding.
In this article, we instead demonstrate a learning algorithms that directly output precodings with no constraint to a codebook.

The predominant choices for deep \ac{rl} algorithms on continuous action spaces are either on-policy like \ac{trpo}~\cite{schulman2017trustregionpolicyoptimization}, \ac{ppo}~\cite{schulman2017proximalpolicyoptimizationalgorithms}, or off-policy like \ac{ddpg}~\cite{lillicrap2019continuouscontroldeepreinforcement}, \ac{sac}~\cite{haarnoja2018soft}.
These algorithms differ primarily not in the performance they achieve, but in how they integrate with a system.
In satellite communications, we are 1)~limited by satellite hardware, 2)~want to limit communication overhead, and 3)~want to limit unnecessary trial-and-error.
Hence, the ability to learn more efficiently on ground-based computation clusters and sample efficiency are main concerns.

On-policy methods such as \ac{ppo} and its variants learn stably but discard samples after only a few gradient updates, which is less efficient when generating new samples is costly. 
Off-policy methods, by contrast, can reuse past experiences more effectively. 
Among them, \ac{ddpg} is a computationally lighter predecessor to \ac{sac} but tends to exhibit much slower convergence and severely lower sample efficiency~\cite{steffenddpg}. 
Among other improvements, the \ac{sac} algorithm augments off-policy learning with a maximum-entropy objective to encourage more stable exploration and improved sample efficiency. 
This makes \ac{sac} particularly well suited for our offline, ground-based training setup, where computational demands are acceptable but data efficiency is essential.
We therefore select \ac{sac} and adapt it to learn precoding strategies for satellite networks.
In our considered scenario, satellites collect precoding data and forward it to a ground station periodically. The ground station updates the parametrized precoding policy using a replay buffer and periodically transmits updated parameters back to the satellites.


The contributions of this work are thus:
\begin{itemize}
	\item We adapt the vanilla \ac{sac} \ac{ml} algorithm to downlink satellite \ac{sdma} precoding, for single satellite, cooperative multi-satellite with local \ac{csit}, and cooperative multi-satellite with global \ac{csit}. Using this, we directly output highly flexible precoding algorithms that maximize an arbitrary objective.
	\item We numerically evaluate learned algorithms that output high sum rate under erroneous \ac{csit}, a NP-hard problem~\cite{roper2023robust}, extending first results presented in earlier works~\cite{gracla_2023_learning, schroeder_2024_flexible}. We compare to analytical baseline precoders on simulated data, considering several combinations of satellite and user constellations as well as error models.
	\item We give particular detail on the \ac{ml}-specific challenges in terms of convergence and learning sample design.
    \item We find that the learned precoders well adapt to diverse challenges in terms of availability and truthfulness of \ac{csit}. Learning leads to high performance in all evaluated contexts while offering significantly increased flexibility over analytical approaches.
\end{itemize}

The paper is structured as follows.
After this introduction, we establish the satellite downlink model, including satellite and user positioning as well as the wireless communications model. 
We then describe the models of uncertainty applied in this work and define what levels of information are available in distributed precoding. 
Finally, we state the sum rate maximization problem.

Following that, \refsec{sec:OURAPPROACH} introduces the use of \ac{rl} in order to find a precoding algorithm.
We step through the learning and inference processes in detail and describe the adaptations made to the baseline \ac{sac} \ac{rl} algorithm.
In \refsec{sec:baselineprecoders}, we define the analytical baseline precoders.
They are used to gauge the learned precoders' performance, but also to show the upsides and potential downsides of analytically derived precoders that promote the use of a data-driven approach.
Finally, \refsec{sec:experiments} evaluates the precoders over a wide variety of scenarios and gives practical insight on the different mechanisms that promote robustness to uncertainty, followed by conclusions.

We assume prior knowledge of basic \ac{nn} function and wireless communication theory.
\reftab{tab:acronyms} lists all abbreviations used.

\textit{Notations:} In the following, we denote
\begin{itemize}
	\item vectors in boldface lowercase letters: \( \mathbf{a} \);
	\item matrices in boldface uppercase letters: \( \mathbf{A} \);
	\item \( \mathcal{U}, \mathcal{N} \) as continuous uniform and Normal distributions;
	\item sets as other uppercase calligraphic letters: \( \mathcal{A} \);
	\item \( |\cdot| \) and \( \|\cdot\| \) as L1 and L2 norms;
	\item \( \circ \) as Hadamard product;
	\item and \( \{ \cdot \}^{\text{H}} \) as the Hermitian of a matrix.
\end{itemize}
\reftab{tab:variables} lists a key selection of denotations.

\begin{table}[!t]
	\renewcommand{\arraystretch}{1.3}%
	\caption{List of Abbreviations}%
	\label{tab:acronyms}%
	\centering%
	\setlength{\tabcolsep}{3pt}%
	\rowcolors{2}{white}{uniblue1!10}%
	\begin{tabular}{p{30pt}p{203pt}}
	\hline
	\acs{acnn} & \acl{acnn}\\
	\acs{aod} & \acl{aod}\\
	\acs{crnn} & \acl{crnn}\\
	\acs{csit} & \acl{csit}\\
	\acs{dl} & \acl{dl}\\
	\acs{iui} & \acl{iui}\\
	\acs{leo} & \acl{leo}\\
	\acs{los} & \acl{los}\\
	\acs{lr} & \acl{lr}\\
	\acs{mimo} & \acl{mimo}\\
	\acs{ml} & \acl{ml}\\
	\acs{mmse} & \acl{mmse}\\
	\acs{nn} & \acl{nn}\\
	\acs{ntn} & \acl{ntn}\\
	\acs{qos} & \acl{qos}\\
	\acs{rl} & \acl{rl}\\
	\acs{sac} & \acl{sac}\\
	\acs{sdma} & \acl{sdma}\\
	\acs{sgd} & \acl{sgd}\\
	\acs{sinr} & \acl{sinr}\\
	\acs{slnr} & \acl{slnr}\\
	\acs{ula} & \acl{ula}\\
	\hline
\end{tabular}%
\end{table}

\begin{table}[!t]
	\renewcommand{\arraystretch}{1.3}%
	\caption{List of Select Denotations}%
	\label{tab:variables}%
	\centering%
	\setlength{\tabcolsep}{3pt}%
	\rowcolors{2}{white}{uniblue1!10}%
	\begin{tabular}{lll}
	\hline
	\emph{Variable} & \emph{Meaning} & \(f(\timeindex)\) \\
	\( \timeindex \) & Simulation step & Yes \\
	\( \stepsperupdate \) & Training synchronization interval & No \\
	\( \satelliteset, \numsats, \satelliteindex \) & Set, number, index related to satellites & No \\
	\( \userset, \numusers, \userid \) & Set, number, index related to users & No \\
	\( \numantennaspersat, \antennaid \) & Number, index related to a satellite's antennas & No \\
	\( \steeringvec, \steeringsca \) & Steering vector, steering vector entry & Yes \\
	\( \distance_{\satelliteset}, \distance_{\userset} \) & Average inter-satellite, inter-user distances & No \\
    \( \mathbf{\txsymbol}, \txsymbol \) & Transmit data vector, data symbol & Yes \\
	\( \mathbf{\txdata}, \txdata \) & Precoded data vector, data symbol & Yes \\
	\( \precodingmatrix, \precodingvector, \precodinginternal \) & Precoding matrix, vector, entry & Yes \\
	\( \csimatrix, \csivector \) & \acs{csit} & Yes \\
	\( \csimatrixestim, \csivectorestim \) & Estimated \acs{csit} & Yes \\
	\( \csimatrixestim_{\satelliteindex}, \csimatrixestimdecone \) & Local estimated \acrshort{csit} & Yes \\
	\( \erroraod, \errorphase \) & Error model 1, 2 realizations & Yes \\
	\( \erroraodbound, \errorphasevariance \) & Error model 1, 2 parametrization & No \\
	\( \sumrate, \ergodicsumrate\) & Sum rate, mean sum rate & Yes \\
	\( \actorfunction, \criticfunction \) & Parametrized actor, critic \acs{nn} functions & on \(\stepsperupdate\) \\
	\( \paramsactor, \paramscritic \) & Actor, Critic \acs{nn} parameters & on \( \stepsperupdate \) \\
	\( \statevec, \actoroutputsampled \) & \acs{rl} precoder input state, output action & Yes \\
	\( \batchset, \batchsize, \batchindex \) & Batch set, size, index & on \(\stepsperupdate\) \\
	\( \loss \) & Loss function & on \( \stepsperupdate \) \\
	\hline
\end{tabular}%
\end{table}



%


\section{THE SATELLITE DOWNLINK MODEL}
\label{sec:setup}

In this section, we first introduce the system setup, followed by the wireless communication model, including the \ac{los} characteristics of the transmission and the signal design with precoding. We then discuss two uncertainty models for \ac{csit}. One reflecting imperfect user position knowledge at the satellites, and another modeling satellite-specific phase errors, for instance due to synchronization mismatches. Building on this, we describe different models of locally available \ac{csit} for cooperative beamforming. Here, we assume perfect synchronization between satellites, but consider that each satellite may only know its own \ac{csit} while having limited or outdated information about others. Finally, we formulate the overall optimization objective, which is to maximize the achievable sum rate.

\subsection{System Setup}

\begin{figure}[!t]
	\centering%
	\includegraphics{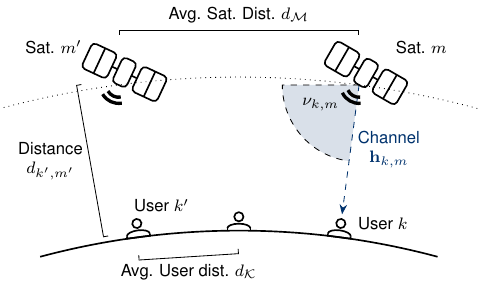}
	\caption{%
		The multi satellite downlink scenario.
		One or more satellites~\( {\satelliteindex \in \satelliteset} \) are performing cooperative downlink transmissions to multiple users~\( {\userid \in \userset} \).
		The \ac{csit}~\( {\csivector_{\userid, \satelliteindex} \in \numberscomplex^{\num{1}\times\numantennaspersat} }\) between satellite~\( \satelliteindex \) and user~\( \userid \) will influence the quality of transmission.
		The satellites make use of estimated, potentially erroneous \ac{csit} to cooperatively form transmission beams.
	}
	\label{fig:scenario}
\end{figure}

We perform downlink transmissions from a set \( { \satelliteset = \{ 1, \ldots, \numsats \} } \) of \ac{leo} satellites at an altitude \( \distance_{0} \) working cooperatively to a set \( { \userset = \{ 1, \ldots, \numusers \} } \) users on Earth, as depicted in \reffig{fig:scenario}.
For simplicity of demonstration, we restrict ourselves to a two dimensional geometric setup, though the principles shown in subsequent sections are not restricted to the two dimensional case.
Each satellite~\( { \satelliteindex\in\satelliteset } \) is equipped with a \ac{ula} of \( {\numantennaspersat} \)~antennas with combined transmit gain~\( \satgain \), and we assume \( { \numsats\numantennaspersat \geq \numusers } \) unless stated otherwise.
Each user~\({\userid \in \userset}\) has a handheld receiver with only a single receive antenna with low receive gain~\( \usergain \), e.g., a mobile phone.
Since the antenna elements of each satellite's \ac{ula} are spatially separated, the output of each antenna element~\( \antennaid \in \{1, \ldots, \numantennaspersat\} \) experiences a different phase shift on its way to the receiver.
Hence, a steering vector \({ \steeringvec_{\userid, \satelliteindex} \in \numberscomplex^{1\times\numantennaspersat} }\) describes the phase shift introduced in each antenna element relative to the geometric center of the \ac{ula}.
Its entries~\( \steeringsca_{\userid, \satelliteindex, \antennaid} \) depend on 1)~the \ac{aod}~\( \aod_{\userid, \satelliteindex} \) between satellite~\(\satelliteindex\) and user~\( \userid \) as well as 2)~the inter-antenna distance \( {\distance_{\antennaid, \antennaid'}} \) of two antennas \( \antennaid, \antennaid' \).
Recall that a \ac{ula} has constant inter-antenna distance, thus, \({ \distance_{\antennaid, \antennaid'} := \distance_{\antennaid} \, \forall \, \antennaid, \antennaid'} \).
To ensure uniform antenna distribution around the center of the array, the steering vector entries for any antenna~\( \antennaid \) of satellite \( \satelliteindex \) calculate as~\cite{tse2005fundamentals}
\begin{equation}
	\label{eq:steeringvectorentry}
	\begin{split}
		&\steeringsca_{\userid, \satelliteindex, \antennaid}(\cos(\aod_{\userid, \satelliteindex})) =\\
		&\exp\left(
		-j \num{2} \pi \frac{\distance_{\antennaid}}{\lambda}
		\frac{\numantennaspersat + \num{1} - \num{2}\antennaid}{2}
		\cos(\aod_{\userid, \satelliteindex})
		\right)
		,
	\end{split}
\end{equation}
with wavelength~\( \wavelength \).

We do not consider a continuous flyover.
Instead, upon a new simulation step~\( {\timeindex \in \numbersnatural } \), all random variables are reevaluated and dependent parameters, including the positions of all satellites~\( \satelliteset \) and users~\( \userset \), are updated. This applies to both the learning phases and the subsequent inference, where the mean performance of a scenario is evaluated using the Monte Carlo method.
For the duration of a simulation step, all parameters are considered as constant, hence, we omit the time index~\( \timeindex \) for brevity. 
In \reftab{tab:variables}, we provide an overview of the parameters that change between different simulation steps~\( \timeindex \), where each step corresponds to a new channel realization and precoding instance.
Satellite and user positioning, used both during learning and inference, are defined by an average inter-satellite distance~\( {\distance_{\satelliteindex, \satelliteindex'} \in \numbersreal^+ }\), average inter-user distance~\( {\distance_{\userid, \userid'} \in \numbersreal^+}\) and movement ranges~\( {\wigglesat, \wiggleusr \in \numbersreal} \) around their initial positions.  
The average distances are assumed as constant for all~\( { \satelliteindex, \satelliteindex' \in \satelliteset, \satelliteindex\neq\satelliteindex' } \) and \( { \userid, \userid' \in \userset, \userid\neq\userid' } \), so we denote \( {\distance_{\userid, \userid'} = \distance_{\userset}},\, {\distance_{\satelliteindex, \satelliteindex'} = \distance_{\satelliteset}} \) for brevity.
Satellites and users are moved from their initial positions by a value drawn from a continuous Uniform distribution~\( \mathcal{U}_{[-\wigglesat,\wigglesat]} \) and \( \mathcal{U}_{[-\wiggleusr,\wiggleusr]} \) for satellites and users respectively.
Thus, for a given scenario, satellites and users have constant average relative positions for \( \timeindex\rightarrow\infty \), but specific realizations can vary drastically depending on the choices of movement ranges~\( \wigglesat, \wiggleusr \).  
\subsection{Wireless Communications Model}
One or multiple \ac{leo} satellites receive complex data symbols \( {\txsymbol_{\userid} \in \numberscomplex} \) for each user~\(\userid \in \userset\) from terrestrial data servers. Since we evaluate our results in terms of achievable rate, the data symbols are assumed to be Gaussian distributed.
The \ac{leo} satellites then further transmit these data symbols to the users~\( \userset \) in downlink using their \acp{ula}, assuming perfect synchronization among the satellites.
We assume a full buffer, i.e., data symbols \( \txsymbol_{\userid} \) for all users~\( \userid  \) are available in every simulation step~\( \timeindex \).
The satellites' \acp{ula} and the receiving users single antennas span a multi-user \ac{mimo} broadcast channel~\cite{tse2005fundamentals}.
As such, to transmit the data symbols \( \txsymbol_{\userid} \), each symbol is spread across the transmit antennas using a digital precoding vector \( {\precodingvector_{\userid} \in \numberscomplex^{\numsats\numantennaspersat\times1}} \), which will be discussed subsequently. All spread symbols are then superimposed to form the transmit signal

\begin{align}
	\mathbf{\txdata} = \sum_{\userid\in\userset} \precodingvector_{\userid}\txsymbol_{\userid},
	\quad \mathbf{\txdata} \in \numberscomplex^{\numsats\numantennaspersat\times 1}
	.
\end{align}
During transmission, we assume a linear channel where the weighted symbols \( \mathbf{\txdata} \) are distorted by the channel conditions between the satellites' antennas and the user antenna, modeled as a channel vector~\({ \csivector_{\userid} \in \numberscomplex^{\num{1}\times\numsats\numantennaspersat} }\), as well as affected by additive complex Gaussian noise~\( { \noise_{\userid} \sim \mathcal{CN}(0, \noisepower) } \).
Thus, the receive signal~\( \rxdata_{\userid} \) at user~\( \userid \) is modeled as
\begin{align}
	\label{eq:receivesignal}
    \rxdata_{\userid} &=
    	\csivector_{\userid}\mathbf{\txdata} + \noise_{\userid} \\
   	&=
    	\underbrace{%
        	\csivector_{\userid} \precodingvector_{\userid} \txsymbol_{\userid}
        }_{\mathclap{\text{User Rx Symbol}}}
        +
        \underbrace{
	        \csivector_{\userid} \sum_{\mathclap{\substack{\userid' \in \userset\\\userid' \neq \userid}}}
	        	\precodingvector_{\userid'} \txsymbol_{\userid'}
	    }_{\mathclap{\text{\acf{iui}}}}
        + \noise_{\userid}
    ,
\end{align}
with \({ \csivector_{\userid} = [\csivector_{\userid, 1}, \ldots, \csivector_{\userid, \numsats}] }\).
We model the channel~\( { \csivector_{\userid, \satelliteindex} \in \numberscomplex^{\num{1}\times\numantennaspersat} } \) between user~\( \userid \) and a satellite~\( \satelliteindex \) as a \ac{los} channel~\cite{you2020massive} with 1)~amplitude fading caused by free space path loss and random large scale fading~\( { \largescalefading\sim \text{Lognormal}(0, \largescalesigma)} \), 2)~phase rotation due to satellite-user distance with phase shift~\( {\phaseshift_{\userid, \satelliteindex}\in [ \num{0}, \num{2}\pi)} \) and 3)~relative phase offsets due to the spatial geometry of the \ac{ula}, collected in the previously introduced steering vector~\( {\steeringvec_{\userid, \satelliteindex} } \).
Accordingly, we define each channel vector
\begin{equation}
	\label{eq:csivector}
	\begin{split}
		&\csivector_{\userid, \satelliteindex} =  \\
		&\underbrace{
			\frac{\wavelength\sqrt{\satgain\usergain}}{\num{4} \pi \distance_{\userid, \satelliteindex}}
			\frac{1}{\sqrt{\largescalefading}}
		}_{\text{Amplitude Fading}}
		\underbrace{
			\exp(-j \phaseshift_{\userid, \satelliteindex})
		}_{\text{Phase Rotation}}
		\underbrace{
			\steeringvec_{\userid, \satelliteindex}(\cos(\aod_{\userid, \satelliteindex}))
		}_{\text{Steering Vector}}
		.
	\end{split}
\end{equation}
For the amplitude fading term, no steering correction is applied as \( \distance_{\antennaid} \ll \distance_{\userid, \satelliteindex} \). We assume the channel vector to be constant for the duration of a simulation instance~\( \timeindex \).

We see in \refeq{eq:receivesignal} that the transmitting satellites can use knowledge of \ac{csit} \( { \csimatrix = [ \csivector_{\num{1}}^{\text{T}}, \ldots, \csivector_{\numusers}^{\text{T}} ]^{\text{T}} }, {\csimatrix\in\numberscomplex^{\numusers\times\numsats\numantennaspersat}} \) (or estimates thereof) to identify a matching digital precoding~\( {\precodingmatrix = [ \precodingvector_{\num{1}}, \ldots, \precodingvector_{\numusers} ]}, { \precodingmatrix\in\numberscomplex^{\numsats\numantennaspersat\times\numusers} } \) that ideally mitigates channel influence \( \csivector_{\userid} \precodingvector_{\userid} \) and suppresses \ac{iui} \(  \csivector_{\userid} \precodingvector_{\userid'} \). The design of suitable precoding methods for perfect \ac{csit} is well studied from terrestrial multi-user \ac{mimo}, e.g., \cite{tse2005fundamentals, marzetta2016fundamentals}, though the findings do not transfer perfectly to \ac{ntn}~\cite{perez2019signal}. 

However, in practice, perfect \ac{csit} cannot be guaranteed, which is why we consider imperfect \ac{csit} according to \refsec{sec:uncertainty} arising from inaccuracies in user position or mismatches in synchronization between satellites. Imperfect \ac{csit} can significantly degrade the precoder performance~\cite{roper2023robust}. 
We present two baseline analytical precoder designs later in \refsec{sec:baselineprecoders}. While one of these baselines is specifically designed to handle \ac{csit} errors, its analytical derivation constrains it to predefined scenarios and limits its adaptability. In contrast, our proposed approach employs \ac{rl}, which directly takes the available \ac{csit}, perfect or imperfect, as input and can flexibly adapt the precoding strategy to the given scenario.

\subsection{Uncertainty Models}
\label{sec:uncertainty}

In traditional precoding approaches, the acquisition of accurate \ac{csit} is a prerequisite. While uplink-based \ac{csi} estimation can suffer from rapid channel aging, \ac{ntn} benefit from predominantly \ac{los} propagation, allowing \ac{csit} to be inferred from user positions and the known satellite antenna geometry. Such position-based estimation is inherently less susceptible to channel aging, as position and geometry vary on much slower time scales than the small-scale fading effects impacting uplink-based estimation.
Nevertheless, various factors can still impair the accuracy of the estimated \ac{csit}. In \ac{leo} systems, the most relevant are the fast relative motion of satellites and users, and, particularly in cooperative multi-satellite beamforming, the latency in distributing \ac{csit} and synchronization among satellites.

In this paper, we apply two different error sources to the \ac{csit} estimate provided to the precoders.

Firstly, as seen in \refeq{eq:csivector}, the \ac{aod}~\( \aod_{\userid, \satelliteindex} \) from satellite~\( \satelliteindex \) toward a user~\( \userid \) is a key contributor to the channel phase.
Thus, we can model a positioning mismatch by introducing an error on the steering vector~\cite{roper2023robust}.
We define a steering error vector
\begin{align}
	\label{eq:errormodel1}
	\errortermaod = \steeringvec_{\userid, \satelliteindex}(\erroraod) \in \numberscomplex^{\num{1}\times\numantennaspersat},
\end{align}
with entries analogous to \refeq{eq:steeringvectorentry}, where \( {\erroraod \sim \mathcal{U}(-\Delta\error_{\text{aod}, \userid, \satelliteindex}, \Delta\error_{\text{aod}, \userid, \satelliteindex}) \in \numbersreal } \) is an error realization from a Uniform distribution.
Given the relative distances, we assume that an error realization~\( \erroraod \) is constant for all antennas of a satellite and assume the same error bound~\( { \Delta\error_{\text{aod}, \userid, \satelliteindex} = \erroraodbound \, \forall \, \userid \in \userset, \satelliteindex \in \satelliteset } \) for all users and satellites.
Note that this error model still leads to different scaling of an error realization on different antennas of the same array.
Effectively, element-wise multiplication of this error vector to a \ac{csit}~\( \csivector_{\userid, \satelliteindex} \) translates to introducing an additive error on \( \cos(\aod_{\userid, \satelliteindex}) \).

As a second error source, we model a total phase error
\begin{align}
	\errortermphase = \exp(-j\errorphase) \in \numberscomplex
\end{align}
with \( {\errorphase \sim \mathcal{N}(0, \errorphasevariance)} \) drawn from a Normal distribution for each satellite~\( \satelliteindex \) and user~\( \userid \).
Such a total phase error per satellite may be caused by a synchronization mismatch.
Given both errors, the \ac{csit} estimate that a satellite~\( \satelliteindex \) may receive for the channel to a user~\( \userid \) is as follows
\begin{align}
	\label{eq:csiestimate}
	\csivectorestim_{\userid, \satelliteindex} = 
		\errortermphase
		\csivector_{\userid, \satelliteindex} \circ
		\errortermaod
\end{align}
with \( \csivectorestim_{\userid}, \csimatrixestim \) defined analogously to the true \ac{csit}.
Given no error, the estimate is perfect, i.e., \( \csimatrixestim|_{\erroraodbound = \errorphasevariance = 0} = \csimatrix \).

So far, we have considered global \ac{csit} for precoding, i.e., all satellites have access to the same complete \ac{csit} matrix, whether perfect \(\csimatrix\) or imperfect \(\csimatrixestim\). Since this assumption may not hold in practical systems due to factors such as communication overhead and delays, we subsequently introduce and investigate distributed precoding based on the local \ac{csit} available at each satellite.

\subsection{Local Information Models}
\label{sec:decentralizedsystem}

In addition to precoding based on global information, we also consider distributed precoding using \emph{local} \ac{csit} at each satellite. In a cooperative precoding scenario without global \ac{csit}, other satellites' information at each satellite may either not be available or have a higher level of uncertainty due to, e.g., being outdated by transmission delay.
We consider three variants of locally available \ac{csit}:

\begin{enumerate}
	\item \emph{Local Precoding}, where, at each inference step~\( \timeindex \), each satellite~\( \satelliteindex \) has access only to its own \ac{csit} estimate and possesses no knowledge of the \ac{csit} of any other satellite \(\satelliteindex'\)
	\begin{align}
		\label{eq:csiblind}
		 \csimatrixestim_{\satelliteindex} = [\csivectorestim_{1, \satelliteindex}^{\text{T}}, \ldots, \csivectorestim_{\numusers, \satelliteindex}^{\text{T}}]^{\text{T}},
		 \quad
		 \csimatrixestim_{\satelliteindex} \in \numberscomplex^{\numusers\times\numantennaspersat}
		 \ .
	\end{align}
    The overall estimated channel can be written as~\(\csimatrixestim = [\csimatrixestim_1 \; \dots \; \csimatrixestim_\numsats]\). Each satellite then computes its own slice of the full precoding matrix 
    \(\precodingmatrix_{\satelliteindex} \in \mathbb{C}^{\numantennaspersat \times \numusers}\) based on its local estimate \(\csimatrixestim_\satelliteindex\),
    such that the global precoding matrix can be assembled as~\( {\precodingmatrix = [ \precodingmatrix^\text{T}_{\num{1}}, \ldots, \precodingmatrix^\text{T}_{\numsats} ]^\text{T}} \).

	\item \emph{Limited 1}, where each satellite~\( \satelliteindex \) has perfect knowledge of their own \ac{csit}~\( {\csimatrix_{\satelliteindex} \in \numberscomplex^{\numusers\times\numantennaspersat}} \) but erroneous \ac{csit}~\( {\csimatrixestim_{\satelliteindex'}\in\numberscomplex^{\numusers\times\numantennaspersat}} \) from other satellites~\( \satelliteindex' \), i.e., satellite~\( \satelliteindex \) performs the precoding based on a channel matrix
	\begin{align}
		\label{eq:csilimited1}
		&\csimatrixestimdecone = [\csimatrixestim_{\satelliteindex'<\satelliteindex}, \csimatrix_{\satelliteindex}, \csimatrixestim_{\satelliteindex'>\satelliteindex}]
		,
		\\
		\nonumber
		&\csimatrixestimdecone \in \numberscomplex^{\numusers\times\numsats\numantennaspersat}
	\end{align}
	\item \emph{Limited 2}, where each satellite has an erroneous estimate of their own \ac{csit} and a further degraded estimate of all other satellites' \ac{csit}. 
	Specifically, we define an erroneous \ac{csit}~\( {\csimatrixestimsat_{\satelliteindex} \in \numberscomplex^{\numusers\times\numantennaspersat}} \) exactly like \( \csimatrixestim_{\satelliteindex} \), except that the error realizations~\(\erroraod\) are multiplied by, for example, two.
	Consequently, for precoding, each satellite has access to a channel matrix
	\begin{align}
		\label{eq:csilimited2}
		&\csimatrixestimdectwo = [
		\csimatrixestimsat_{\satelliteindex'<\satelliteindex},
		\csimatrixestim_{\satelliteindex},
		\csimatrixestimsat_{\satelliteindex'>\satelliteindex}]
		,
		\\
		\nonumber
		&\csimatrixestimdectwo \in \numberscomplex^{\numusers\times\numsats\numantennaspersat}
		.
	\end{align}
\end{enumerate}

\subsection{Problem Statement}
\label{sec:problem}

In this paper, the measure of a precoding~\( { \precodingmatrix } \) is the achieved sum rate
\begin{align}
\label{eq:sumrate}
	\sumrate = 
		\sum_{\userid\in\userset}
		\log\left(\num{1} +
		\frac{\left| \csivector_{\userid} \precodingvector_{\userid} \right|}
		{\noisepower + \sum_{\userid' \in \userset, \userid' \neq \userid} \left| \csivector_{\userid}\precodingvector_{\userid'} \right|}
		\right)
	.
\end{align}
Given the stochastic fading channel and uncertain \ac{csit} as defined in the prior sections, we will seek to maximize the mean sum rate
\begin{align}
	\label{eq:ergodicsumrate}
	\ergodicsumrate & =
		\expectation_{\csimatrix}\left[
			\sum_{\userid\in\userset}
			\log\left(\num{1} +
			\frac{\left| \csivector_{\userid} \precodingvector_{\userid} \right|}
			{\noisepower + \sum_{\userid' \in \userset, \userid' \neq \userid} \left| \csivector_{\userid}\precodingvector_{\userid'} \right|}
			\right)
		\right]
		\\
		& =
		\expectation_{\csimatrix}\left[
			\sumrate
		\right]
\end{align}
which we approximate from the average sum rates achieved.
The satellites also share a power budget \( \txpower \in \numbersreal \).
Thus, we formulate the general optimization objective as
\begin{subequations}
	\label{eq:optimizationobjective}
	\begin{alignat}{2}
		&\max_{\precodingmatrix} & \quad
			& \ergodicsumrate 
		\\
		\label{eq:optimizationobjectivepower2}
        &\quad \subjectto & 
			&\|\precodingmatrix_{\satelliteindex}\|^{\num{2}} \leq \txpower / \numsats \quad \forall \, \satelliteindex \in \satelliteset
		.
	\end{alignat}
\end{subequations}
This problem is known to be NP-hard~\cite{roper2023robust}.
To ensure comparable results, we distribute the power budget equally among all satellites according to \refeq{eq:optimizationobjectivepower2}, so that constellations with more \ac{leo} satellites do not inherently benefit from higher total transmit power.
In the case of distributed precoding, instead of optimizing the global precoding matrix~\(\precodingmatrix\) in \refeq{eq:optimizationobjective}, each satellite~\(\satelliteindex\) optimizes their own precoding matrix~\( \precodingmatrix_{\satelliteindex} \). 


%


\section{REINFORCEMENT LEARNED PRECODERS}
\label{sec:OURAPPROACH}
Compared to reality, we can reasonably assume any mathematical representation of both the communications model and error model to be imperfect.
We might also encounter a problem such as the sum rate maximization that is too complex or resource intensive to solve directly, thus working with relaxed targets or substituting approximations.
As an alternative, we may forego deriving a precoding algorithm from the flawed model and instead infer it from given data.
\ac{rl} approaches this with a process of trial-and-error learning, where a learning agent interacts with the environment by selecting a precoding~\( \precodingmatrix \) for a given channel state estimation~\( \csimatrixestim \) and receiving the resulting sum rate~\( \sumrate \).
Based on this experienced interaction, the agent adapts its future behavior to promote decisions that lead to high sum rate~\( \sumrate \) and deemphasize actions that result in lower sum rate~\( \sumrate \).
Thus, using \ac{rl}, we make no model assumptions and require no pre-existing labeled data set.
In typical \ac{rl} parlance, this problem may also be stated as a \ac{mdp}. 
Here, the channel state estimations~\( \csimatrixestim \) are part of the set of possible states as spanned by the system model, the precodings~\( \precodingmatrix \) come from action set of all possible precodings, and the sum rate~\( \sumrate \) corresponds to the reward emitted when transitioning to a new state.
The state transition probability \(\text{Prob}(\csimatrixestim_{\timeindex}, \csimatrixestim_{\timeindex+1})\) is defined by the system model discussed in the previous section.

In terms of selecting a learning algorithm from the family of \ac{rl}, we can assume limitations in compute hardware and power onboard the satellites.
Further, we can assume that the system, i.e., the downlink channels properties, will not structurally change on a short time scale or unexpectedly.
State of the art on-policy \ac{rl} algorithms such as \ac{trpo}~\cite{schulman2017trustregionpolicyoptimization}, \ac{ppo}~\cite{schulman2017proximalpolicyoptimizationalgorithms} quickly discard learning samples once the policy is changed, reducing sample efficiency.
Hence, we suggest that data samples are stored and learning is executed off-policy and offline at a terrestrial data center.
The trained models can then be distributed to the satellites as required.
Should the system change structurally, e.g., through changed user hardware assumptions, pre-trained models can be adapted to new data~\cite{huisman2021survey}, though we do not consider that case in this paper.
Offline training also allows for easy bootstrapping or other augmentation of the data set by simulated data, e.g.,~\cite{nvidia2024replicator}, which we do use in this paper.
Finally, we wish to be sample efficient in selecting training trials, as generating data samples incurs a cost in terms of transmission and, potentially, system performance.
We therefore opt to use the state-of-the-art \ac{sac}~\cite{haarnoja2018soft} learning algorithm, which we describe in detail in the following.
It learns offline from a buffer of data samples and imposes maximum entropy in the Bayesian sense to generate high value data samples.
Subsequent to that, in \refsec{sec:sacadditions}, we will describe some additions to the vanilla \ac{sac} that were necessary to achieve high performance.

\subsection{Soft Actor Critic}
\label{sec:TOPIC1}

\begin{figure}
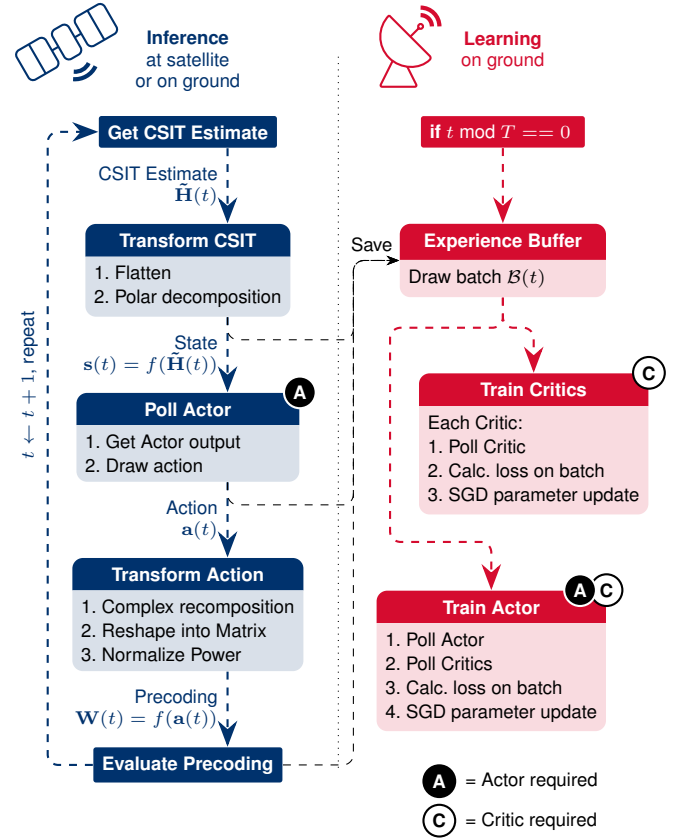

	\centering%
	\pgfmathsetmacro{\xshiftspacing}{15}
\begin{tikzpicture}
[
	scale=1,
	every node/.style={font=\sffamily\scriptsize},
	iblock/.style={
		fill=uniblue1,
		text=white,
		rounded corners=1,
		align=center,
	},
	lblock/.style={
		fill=unired2,
		text=white,
		rounded corners=1,
		align=center,
	},
	iblocksplit/.style={
		rectangle split,
		rectangle split parts=2,
		rectangle split part fill={uniblue1,uniblue1!15},
		rounded corners,
		align=center,
	},
	lblocksplit/.style={
		rectangle split,
		rectangle split parts=2,
		rectangle split part fill={unired2,unired2!15},
		rounded corners,
		align=center,
	},
    criticblocksplit/.style={
		rectangle split,
		rectangle split parts=2,
		rectangle split part fill={unipeach1,unired2!15},
		rounded corners,
		align=center,
	},
    actorblocksplit/.style={
		rectangle split,
		rectangle split parts=2,
		rectangle split part fill={black,unired2!15},
		rounded corners,
		align=center,
	},
    actorbubble/.style={
        circle,
        draw=white,
        thick,
        text=white,
        fill= black,
        minimum size=12pt,
        inner sep=0pt,
        align=center
    },
    criticbubble/.style={
        circle,
        draw=black,
        thick,
        text=black,
        fill= white,
        minimum size=12pt,
        inner sep=0pt,
        align=center
    },
	a/.style={
		-{Stealth[scale=1.5]},
		dashed,
        thick,
		rounded corners,
	},
	ia/.style={
		a,
		color=uniblue1,
	},
	la/.style={
		a,
		color=unired2,
	},
	asmol/.style={
		-{Stealth},
		dashed,
		rounded corners,
	}
]

	\input{figures/satellite.sty}
	\input{figures/satellitedish.sty}

	\satellite{(-1.5, 1)}{30}{0.5}{color=uniblue1}
	\satellitedish{(2.8, 0.6)}{0}{0.5}{color=unired2}

	
	\node (igetcsi)
		[iblock]
		{\bfseries Get CSIT Estimate};
		
	\node (itransformcsi)
		[iblocksplit, below = of igetcsi]
		{%
			\color{white}
			\bfseries
			Transform CSIT
			\nodepart[align=left]{two}
			1. Flatten
			\\[.3ex]
			2. Polar decomposition
		};
		
	\node (ipollactor)
		[iblocksplit, below = of itransformcsi]
		{%
			\color{white}
			\bfseries
			Poll Actor
			\nodepart[align=left]{two}
			1. Get Actor output
			\\[.3ex]
			2. Draw action
			\phantom{aaaaaaaa}
		};
		
	\node (itransformaction)
		[iblocksplit, below = of ipollactor]
		{%
			\color{white}
			\bfseries
			Transform Action
			\nodepart[align=left]{two}
			1. Complex recomposition
			\\[.3ex]
			2. Reshape into Matrix
			\\[.3ex]
			3. Normalize Power
		};
		
	\node (ievaluate)
		[iblock, below = of itransformaction]
		{\bfseries Evaluate Precoding};
		
	
	\node (lcriterion)
		[lblock, right = 1.9 of igetcsi]
		{%
			\textbf{if} \( \timeindex \text{ mod } \stepsperupdate == 0 \)
		};
		
	\node (ldrawbatch)
		[lblocksplit, below = of lcriterion]
		{%
			\color{white}
			\bfseries
			Experience Buffer
			\nodepart[align=left]{two}
			Draw batch \({ \batchset(\timeindex) }\)
			\phantom{aaaa}
		};
	
	\node (ltraincritic)
		[lblocksplit, below right = 1 and -2.5 of ldrawbatch]
		{%
			\color{white}
			\bfseries
			Train Critics
			\nodepart[align=left]{two}
			Each Critic:
			\\[.3ex]
			1. Poll Critic
			\\[.3ex]
			2. Calc. loss on batch
			\\[.3ex]
			3. SGD parameter update
		};

	\node (ltrainactor)
		[lblocksplit, below left = 1 and -2.5 of ltraincritic]
		{%
			\color{white}
			\bfseries
			Train Actor
			\nodepart[align=left]{two}
			1. Poll Actor
			\\[.3ex]
			2. Poll Critics
			\\[.3ex]
			3. Calc. loss on batch
			\\[.3ex]
			4. SGD parameter update
		};

    \node (actorbubblepoll) at  (ipollactor.north east) 
        [actorbubble]    
        {\bfseries A};

    \node (criticbubblecritic) at  (ltraincritic.north east) 
        [criticbubble]    
        {\bfseries C};

    \node (criticbubbleactor) at  (ltrainactor.north east) 
        [criticbubble]    
        {\bfseries C};
    \node (actorbubbleactor) at  ([xshift=-10pt]ltrainactor.north east) 
        [actorbubble]    
        {\bfseries A};

	\draw [ia]
		([xshift=\xshiftspacing]igetcsi.south)
		--
		node [left, align=right] {\acs{csit} Estimate \\ \( \csimatrixestim(\timeindex)\)}
		([xshift=\xshiftspacing]itransformcsi.north);
		
	\draw [ia]
		([xshift=\xshiftspacing]itransformcsi.south)
		--
		node [left, align=right] { State\\  \( \statevec(\timeindex) = \genericfunction (\csimatrixestim(\timeindex)) \)}
		([xshift=\xshiftspacing]ipollactor.north);
		
	\draw [ia]
		([xshift=\xshiftspacing]ipollactor.south)
		--
		node [left, align=right] {Action\\\( \actoroutputsampled(\timeindex) \)}
		([xshift=\xshiftspacing]itransformaction.north);
		
	\draw [ia]
		([xshift=\xshiftspacing]itransformaction.south)
		--
		node [left, align=right] {Precoding\\  \( \precodingmatrix(\timeindex) = \genericfunction(\actoroutputsampled(\timeindex)) \)}
		([xshift=\xshiftspacing]ievaluate.north);
		
	\draw [ia]
		(ievaluate.west)
		--
		++(-0.6, 0)
		|- node [rotate=90, above, pos=0.3] {\(\timeindex \leftarrow \timeindex+1\), repeat}
		(igetcsi.west);
		
	\draw [la]
		(lcriterion)
		--
		(ldrawbatch);
	
	\draw [la]
		(ldrawbatch.south)
		--
		++(0, -0.4)
		-|
		(ltraincritic.north);

	\draw [la]
		(ldrawbatch.south)
		--
		++(0, -0.4)
		--
		++(-1.5, 0)
		--
		++(0, -2.85)
		-|
		(ltrainactor.north);
		
	\draw [asmol]
		([xshift=\xshiftspacing]itransformcsi.south)
		--
		++(0, -0.3)
		-|
		(2.2, -2.1)
		|- node [above, pos=0.7] {Save}
		(ldrawbatch.west);
		
	\draw [asmol]
		([xshift=\xshiftspacing]ipollactor.south)
		--
		++(0, -0.3)
		-|
		(2.2, -2)
		|-
		(ldrawbatch.west);
		
	\draw [asmol]
		(ievaluate.east)
		-|
		(2.2, -2)
		|-
		(ldrawbatch.west);
	
		
	\draw [dotted]
		(2, 1)
		--
		(2, -8.5)
		;
		
	\node (iheader)
		[above = 1.25 of igetcsi, anchor = north, color=uniblue1, align=center]
		{\textbf{Inference}\\ at satellite \\or on ground
        };
		
	\node (lheader)
		[above = 1.25 of lcriterion, anchor = north, color=unired2, align=center]
		{\textbf{Learning}\\ on ground};
		
	\node (actorbubblelegend) 
        [actorbubble, below left = 0.5 and -1 of ltrainactor, draw=black]    
        {\bfseries A};
    \node (actorbubbledescription)
        [right = 0 of actorbubblelegend]
        {= Actor required};

    \node (criticbubblelegend) 
        [criticbubble, below = 0.075 of actorbubblelegend]    
        {\bfseries C};
    \node (criticbubbledescription)
        [right = 0 of criticbubblelegend]
        {= Critic required};

\end{tikzpicture}
	\caption{%
		Flowchart of the inference loop at a satellite or satellite controller on the left half and a learning step at a ground-based server on the right half.
	}
	\label{fig:sacflowchart}
\end{figure}

In \ac{rl}, broadly, an \emph{agent} interacts with the \emph{environment}, in this case by selecting a precoding~\( \precodingmatrix \) according to~\refsec{sec:setup}.
This will result in a certain measure of success, e.g., a sum rate~\( \sumrate \). The agent will try to learn from this interaction to take more successful actions in the future.
The core of our agent is the \gls{sac} \ac{rl} algorithm.
The \ac{sac}, as we use it in this paper, makes use of three \ac{nn}: the \ac{acnn} estimates precodings~\( \precodingmatrix \) for a given system state.
Two \ac{crnn} estimate the goodness of a precoding~\( \precodingmatrix \) for a given state and guide the \ac{acnn} to make better decisions.
As an offline learning algorithm, \ac{sac} also uses an Experience Buffer to store past interaction samples.
Two modules exist to transform \ac{nn} inputs and outputs to embed with the system.
\reffig{fig:sacflowchart} shows the \ac{sac} process flow, where we see the decoupled inference loop and learning step. 
In the following, we describe the design of the depicted modules in more detail, starting with the input transform.
We then return to \reffig{fig:sacflowchart} to describe a full iteration of inference and learning each.

\textit{Input Transform:} The estimated \ac{csit} \( { \csimatrixestim }\), as \refsec{sec:setup} describes, is the system information that our learned precoder, the \ac{acnn}, will use to decide on a precoding~\( \precodingmatrix \).
We are using real-valued \ac{nn} while the \ac{csit}~\( {\csimatrixestim \in \numberscomplex^{\numusers\times\numsats\numantennaspersat}} \) is complex valued.
Hence, we transform~\( \csimatrixestim \) into a flat, real valued vector~\( { \statevec \in \numbersreal^{\num{1} \times \num{2}\numusers\numsats\numantennaspersat} }\) by decomposing each complex channel entry in~\( \csimatrixestim \) into two real-valued components.
Specifically, for the input transform we select the decomposition into phase and amplitude information, which proves to be inductive to learning convergence.
The order of values does not matter as we will not be using, e.g., convolutional \ac{nn}.

\textit{\acl{acnn}:} The \ac{acnn} is a fully connected \ac{nn}~\cite{Goodfellow-et-al-2016} represented by a function~\( { \actorfunction_{\paramsactor} (\statevec) } \) parametrized by a set of weights~\( \paramsactor \).
For readability in notation we omit the parametrization index in the following.
The \ac{nn} takes a transformed channel state~\( \statevec \) as input, performs sequential nonlinear transforms on it, and outputs a vector~\( { \actorfunction(\statevec) = \actoroutput \in \numbersreal^{\num{1}\times\num{4}\numusers\numsats\numantennaspersat} } \) with four values per entry of a precoding~\( { \precodingmatrix } \).
The output vector~\( \actoroutput \) effectively decomposes into two parts of equal lengths, \( \actoroutput = [\actoroutputmeans, \actoroutputlogstd] \), where \( \actoroutputmeans \) represents the estimated real-valued components of a precoding matrix~\( \precodingmatrix \) and \( \actoroutputlogstd \) is a measure of uncertainty for each entry.
During training, the uncertainty is used to sample a real-valued vector~\( {\actoroutputsampled \sim \mathcal{N}(\actoroutputmeans, \exp(\actoroutputlogstd))}, {\actoroutputsampled \in \numbersreal^{\num{1}\times\num{2}\numusers\numsats\numantennaspersat}} \) from a Normal distribution. We later discuss how this is used as a built-in sample selection mechanism. During inference, we instead simply select~\( \actoroutputsampled = \actoroutputmeans \).
The output transform then converts the sampled network output~\( \actoroutputsampled \) into a real precoding matrix~\( \precodingmatrix \).

\textit{\acl{crnn}:} Like the \ac{acnn}, a \ac{crnn} is a fully connected \ac{nn}. It is represented by a function~\( \criticfunction_{\paramscritic} (\statevec, \actoroutputsampled) \) parametrized by a set of weights~\( \paramscritic \).
As before, we will omit the parameter index from here on for readability.
A \ac{crnn} estimates the expected mean sum rate~\refeq{eq:ergodicsumrate} as \( { \criticfunction (\statevec, \actoroutputsampled) = \hat{\sumrate} \in \numbersreal } \) for a given transformed system state~\( \statevec \) and sampled \ac{acnn} output~\( \actoroutputsampled \), i.e., it has only one output~\( \hat{\sumrate} \).
\acp{crnn} are only used in the learning step, guiding the \ac{acnn}'s learning.
This implementation makes use of two independently initialized, but functionally identical \acp{crnn}~\( {\criticfunction_{\idx}, \idx \in \{1, 2\}} \) for learning stability, as explained in the subsequent learning step section.

\textit{Output Transform:} To transform the sampled \ac{acnn}  output~\( \actoroutputsampled \) into a complex-valued precoding matrix~\( \precodingmatrix \), we separate the sampled output~\( { \actoroutputsampled = [\actoroutputsampledreal, \actoroutputsampledimag] } \) into two vectors~\( {\actoroutputsampledreal, \actoroutputsampledimag \in \numbersreal^{\num{1}\times\numusers\numsats\numantennaspersat}} \) of equal length.
We pair two values from each vector such that \({ \precodinginternal_{\idx} = \actoroutputsampledrealsca + j\actoroutputsampledimagsca \,\forall\, \idx \in \{ \num{1}, \ldots, \numusers\numsats\numantennaspersat \}} \).
We then reshape the~\( \precodinginternal_{\idx} \) to a matrix~\( {\precodingmatrix \in \numberscomplex^{\numusers\times\numsats\numantennaspersat}} \).
The specific choice of reshaping algorithm does not matter as the order is learned implicitly.
Power normalization according to the power budget from \refeq{eq:optimizationobjectivepower2} is applied to the precoding matrices of each satellite, ensuring that all satellites are identical in design.

\textit{Experience Buffer:} Any tuple~\({ \experience = ( \statevec, \actoroutputsampled, \sumrate )} \) of transformed \ac{csit}~\( \statevec \), sampled \ac{acnn} outputs~\( \actoroutputsampled \) and achieved sum rate~\( \sumrate \) is saved in a ring buffer with \( \expbuffersize \)~elements.
They are indexed as \( \experience_{\expbufferindex} \) with \({ \expbufferindex \in \{1, \ldots, \expbuffersize'\} }\) where \( \expbuffersize' \) is the current number of sample tuples in the buffer at simulation step~\( \timeindex \), i.e., \( { \expbuffersize' = \expbuffersize } \) for a full buffer.
As a ring buffer, the indexation~\( \expbufferindex \) and number of elements~\( \expbuffersize' \) may change whenever a new sample is added on each~\( \timeindex \).

\subsection{Inference Loop and Learning Step}
With all elements of the vanilla \ac{sac} learning agent known, we return to \reffig{fig:sacflowchart} where the left, blue half depicts the \emph{inference loop}.
At a time~\( \timeindex \) all users and satellites update their positions and the satellites obtain estimated \ac{csit}~\( \csimatrixestim \) as described in \refsec{sec:setup}.
The estimated \ac{csit} is transformed into the \ac{acnn} input~\( \statevec \), and forwarded through the \ac{acnn} to obtain~\( \actoroutputsampled \).
The output transform reshapes the sampled \ac{acnn} output~\( \actoroutputsampled \) into a matrix~\( \precodingmatrix \).
Finally, the precoding~\( \precodingmatrix \) is evaluated according to \refsec{sec:setup}, resulting in a sum rate~\( \sumrate \).
The simulation step~\( \timeindex \) concludes by adding the obtained tuple~\({ \experience = ( \statevec, \actoroutputsampled, \sumrate )} \) to the experience buffer.
As such, an inference step is fairly light-weight and may be carried out onboard a satellite with dedicated \ac{ml} hardware.

For the \emph{learning step}, we refer to the right, red half in \reffig{fig:sacflowchart}. Once the experience buffer has filled with a minimum number~\( \minsamples \) of learning samples, we begin updating the \ac{acnn} and \ac{crnn} every~\( {\stepsperupdate\in\numbersnatural} \) simulation steps.
Both the \ac{crnn} and \ac{acnn} networks' parameters are updated using \ac{sgd}.
Hence, we select a batch~\( { \batchset = \{ \experience_{\mathbf{\idx}} \}_{\mathbf{\idx} \sim \mathcal{U}[1, \expbuffersize']} } \) of experiences sampled uniformly from the current buffer.
The size of the batch is denoted as~\( { | \batchset | = \batchsize} \).
For the \acp{crnn}~\( \criticfunction \), the aim is to best match the concealed mapping \( {f(\statevec, \actoroutputsampled) = \sumrate} \) of the erroneously estimated transformed channel state~\( \statevec \) and a given sampled \ac{acnn} output to a resulting sum rate~\( \sumrate \).
Hence, for the sampled batch~\( \batchset \) of experienced transitions, the mean square loss is
\begin{align}
	\label{eq:losscritic}
	\loss_{\criticfunction, \idx}(\batchset) =
		\frac{\num{1}}{\batchsize}
		\sum_{(\statevec_{\batchindex}, \actoroutputsampled_{\batchindex}, \sumrate_{\batchindex})\in\batchset}
			(
				\criticfunction_{\idx}(
					\statevec_{\batchindex},
					\actoroutputsampled_{\batchindex}
				)
				-
				\sumrate_{\batchindex}
			)^{\num{2}}
\end{align}
for each \ac{crnn}~\( \idx \in \{1, 2\} \) and update their parameters~\( { \paramscriticnum{\idx}} \) accordingly via \ac{sgd},
\begin{align}
	\label{eq:sgdcritic}
	\paramscriticnum{\idx}(\timeindex) \gets \paramscriticnum{\idx}(\timeindex - 1) - \learningratecritic\nabla_{\paramscriticnum{\idx}}\loss_{\criticfunction, \idx}
	,
\end{align}
with a \ac{lr}~\( \learningratecritic \).

For the \ac{acnn}, we wish to find actor outputs~\( \actoroutputsampled \) that maximize the received sum rate~\( \sumrate \) for a given transformed state~\( \statevec \) after the output transformation.
The mapping \( {f(\statevec, \actoroutputsampled) = \sumrate} \) is unknown and potentially undifferentiable.
As a substitute, we have learned the known and differentiable \ac{crnn} mappings~\( {\criticfunction(\statevec, \actoroutputsampled) = \hat{\sumrate}} \).
It has been shown in, e.g., \cite{fujimoto2018addressing} that misestimating the value of a state tends to be detrimental to the learning progress.
Hence, using multiple, independently initialized \acp{crnn} to guide the \ac{acnn} stabilizes the learning process, especially early on.
Thus, using both \acp{crnn}, the \ac{acnn}'s estimated sum rate loss is first defined as
\begin{align}
	\label{eq:actorlosssumrate}
	\loss_{\actorfunction, 1}(\batchset) = 
		\frac{\num{1}}{\batchsize}
		\sum_{\statevec_{\batchindex}\in\batchset}
			-\min
			\left(
				\criticfunction_{1}(\statevec_{\batchindex}, \actoroutputsampled(\statevec_{\batchindex})),
				\criticfunction_{2}(\statevec_{\batchindex}, \actoroutputsampled(\statevec_{\batchindex}))
			\right)
	.
\end{align}
Learning from this loss most closely corresponds to the optimization objective~\refeq{eq:optimizationobjective}, optimizing for a maximum sum rate~\( \ergodicsumrate \) in the mean sense.
However, \ac{sac} also aims to find the maximum entropy solution using the \ac{acnn}'s uncertainty measures for each output.
Therefore, an \ac{acnn} entropy loss is defined as
\begin{align}
	\label{eq:actorlossentropy}
	\loss_{\actorfunction, 2}(\batchset) =	
		\frac{\num{1}}{\batchsize}
		\sum_{\statevec_{\batchindex}\in\batchset}
		\text{logprob}_{\actoroutputsampled\sim\mathcal{N}(\actoroutputmeans(\statevec_{\batchindex}), \exp(\actoroutputlogstd(\statevec_{\batchindex})))}\left(
			\actoroutputsampled(\statevec_{\batchindex})
		\right)
	,
\end{align}
using the log probabilities of the actions~\( \actoroutputsampled(\statevec_{\batchindex}) \) that were sampled for the training batch states~\( \statevec_{\batchindex} \).
Recall that the actions~\( \actoroutputsampled(\statevec_{\batchindex}) \) are sampled from a distribution~\({\mathcal{N}(\actoroutputmeans(\statevec_{\batchindex}), \exp(\actoroutputlogstd(\statevec_{\batchindex})))}\) parametrized by the \ac{acnn} outputs~\( \actoroutputmeans(\statevec_{\batchindex}), \actoroutputlogstd(\statevec_{\batchindex}) \).
Hence, the entropy loss~\refeq{eq:actorlossentropy} may be minimized in the mean sense by increasing the output variance, i.e., changing the actor parameters to produce higher~\( \actoroutputlogstd \).
By increasing the variance of sampled values, the \ac{acnn} generates more varied data samples.
Vice versa, output values that are known to produce significant returns, i.e., low loss \refeq{eq:actorlosssumrate}, may be sampled with lower variance.
Hence, the two losses~\refeq{eq:actorlosssumrate}, \refeq{eq:actorlossentropy} are combined for the \ac{acnn} loss
\begin{align}
	\label{eq:actorloss}
	\loss_{\actorfunction}(\batchset) = 
		\loss_{\actorfunction, 1}
		+
		\exp(\entropyscale) \loss_{\actorfunction, 2}
\end{align}
with a balancing factor~\( \entropyscale \).
Based on this batch loss~\( \loss_{\actorfunction} \), the \ac{acnn} parameters~\( \paramsactor \) are also updated with an \ac{sgd} update step
\begin{align}
	\label{eq:sgdactor}
	\paramsactor(\timeindex) \gets \paramsactor(\timeindex-1) - \learningrateactor \nabla_{\paramsactor}\loss_{\actorfunction}
	,
\end{align}
at a \ac{lr}~\( \learningrateactor \).
The updated \ac{acnn} parameters are then distributed to the agents.
However, we find that the vanilla \ac{sac}~\cite{haarnoja2018soft} as described above is not reliably able to handle the communications problem statement as described in \refsec{sec:setup}.
Hence, we augment the vanilla algorithm using techniques from the domain of \ac{ml} that we describe in the following.
We then provide algorithm boxes that describe the full workflow.

\subsection{Pitfalls \& Additions}
\label{sec:sacadditions}

Applying \ac{ml} often requires minute adjustments to vanilla learning algorithms that sometimes go unreported.
In the following, we aim to list the less prominent challenges and adjustments enabling vanilla \ac{sac} learning to precode, as well as briefly discuss their reasoning.

\emph{Exploitation of simulation limitations:}
Broadly speaking, \ac{rl} may extract any kind of information available from given data, which may yield unexpected performance gains over analytical approaches that do not use that particular information.
Unfortunately, for synthetic data, this may include extracting patterns that would not be found in real data.
If, e.g., the precoder was learning on a static constellation of user and \ac{leo} satellites, it would implicitly infer the user positions regardless of estimation error and achieve approximately perfect performance always.
This would be useless in real applications.
We have sanitized data generation from exploitable simulation artifacts.
However, we see in the evaluation that the learned precoders utilize model assumptions that are not considered in the design of the benchmark analytical precoders, e.g., relatively lower error on the power fading component of the channel.
This is explicitly a strength of data-driven algorithm design.

\emph{Low frequency network updates:}
With our proposed method of learning precoders offline at a ground based server and distributing the learned precoders to the satellites, it is already intuitive to not update the networks at every inference step, but only every \( \stepsperupdate \)~steps.
Reducing the communication overhead is desirable, and with \ac{leo} satellites, updates may not even possible when there is no communication link between the ground server and the satellite.
Additionally, a lower update frequency has also shown to be beneficial to offline learning. 
\cite{fedus2020revisiting} show that the ratio of learning updates to data samples in buffer is a decisive factor in learning convergence.
With less frequent updates, older, potentially outdated data samples are pushed out of the buffer more quickly while maintaining a large, varied sample data collection.

\emph{Input standardization:}
It is well known that, given normal \ac{sgd} parameter updates, normalizing each individual \ac{nn} input channel to a mean and covariance that is matched to the first layer's selected nonlinearity speeds up training~\cite{lecun2002efficient}.
Shifting the mean removes an update direction bias, and scaling the inputs balances the parameter update size.
Additionally, with two different types of inputs as in our case with phase and magnitude components of~\( \csimatrixestim \), scaling balances the relative importance of each factor.
However, shifting and scaling inputs requires knowledge of the data statistics, which we cannot assume as known.
Using a hypothesis test, we confirm that the scaling statistics for magnitude as well as the phase variance may be determined within a relatively low number of samples, i.e., \num{100} samples achieve within \( \pm \SI{10}{\percent} \) of the true statistics at a significance level of \SI{5}{\percent}.
Since these statistics are unlikely to change rapidly or at all, we consider it realistic to include a warm-up period to sample these statistics at the very beginning of a learning process.
Thereafter, the warm-up does not need to be repeated unless the \ac{nn} need to be retrained from scratch.

\emph{Inter layer standardization via Batch Normalization:}
Same as the \ac{nn} inputs, normalizing the connections of the in-between layers of the \ac{nn} benefits learning speed.
However, unlike the first layer, the statistics of these connections shift constantly as the \ac{nn} parameters are updated.
Batch normalization layers~\cite{ioffe2015batch} maintain a moving average of the statistics sampled from each learning batch~\( \batchset \) and normalize by these.
While approximated normalization will not yield the optimum statistics, the noise introduced by normalizing with a moving average regularizes the parameter updates comparable to the noise introduced by \ac{sgd} batches.
We find alternatingly stacking fully connected layers and batch normalization layers to have a significant positive effect.

\emph{Weight Penalty:}
L2 weight regularization~\cite[Chp. 7]{Goodfellow-et-al-2016} is one of the most basic regularization techniques in \ac{ml}. It introduces losses
\begin{align}
	\label{eq:lossweight}
	\loss_{\paramsactor} &= 
		\frac{\num{1}}{\num{2}} \|\paramsactor\|^{\num2}, \\
	\loss_{\paramscritic} &= 
		\frac{\num{1}}{\num{2}} \|\paramscritic\|^{\num2}
\end{align}
on the magnitude of \ac{acnn} and \ac{crnn} weights~\( \paramsactor, \paramscritic \).
The optimization losses~\refeq{eq:actorloss}, \refeq{eq:losscritic} are regularized to
\begin{align}
	\label{eq:lossactorfinal}
	\loss'_{\actorfunction} &= \loss_{\actorfunction} + \weightregulscaleactor \loss_{\paramsactor}, \\
	\label{eq:losscriticfinal}
	\loss'_{\criticfunction} &= \loss_{\criticfunction} + \weightregulscalecritic \loss_{\paramscritic}
\end{align}
with weights~\( \weightregulscaleactor, \weightregulscalecritic \).
The benefit of this is two-fold:
1)~individual parameters with little influence on an inference are pulled to zero, reducing artifacting;
2)~the growth of parameters from within a single batch~\( \batchset \) of data is limited, preventing overly dominant paths.

With these adjustments in place, the \ac{sac} algorithm is able to reliably learn well performing precoders.
Algorithms~\ref{alg:inference}, \ref{alg:learning} detail a full inference and learning step, respectively.

\subsection{Adjustments for Distributed Precoding}
In distributed scenarios, as described in \refsec{sec:setup}\ref{sec:decentralizedsystem}, each satellite individually determines their own slice~\( \precodingmatrix_{\satelliteindex} \) of the full precoding matrix~\( \precodingmatrix \) given local \ac{csit}.
Hence, to adjust for this scenario, multiple \acp{sac} (one per satellite) are deployed and trained in parallel at the ground based computation cluster.
Once trained, each \ac{nn} is distributed to its corresponding satellite and precodings can be generated on-board.
The \acp{nn} input and output dimensions are adjusted according to the given local information.
No further adjustment is required.

Next, we introduce some analytical precoding baselines before evaluating the all precoders' performance.

\begin{algorithm}[t]
	\caption{Inference}%
	\label{alg:inference}%
	\KwIn{Estim. \acs{csit} \( \csimatrixestim(\timeindex) \)}%
\KwData{Input norm. factors, \acs{acnn}~\( \actorfunction \)}%
\KwResult{Precoding \( \precodingmatrix(\timeindex) \) achieves sum rate \( \sumrate(\timeindex) \)}%

	\(\statevec(\timeindex) \gets \text{InputTransform}(\csimatrixestim(\timeindex))  \)
	
	\( \statevec'(\timeindex) \gets \text{Normalize}(\statevec(\timeindex)) \)
	
	\( \actoroutput(\timeindex) \gets \actorfunction(\statevec'(\timeindex)) \) 
	
	\eIf{GenerateTrainingSamples}
	{	
		\( \actoroutputsampled(\timeindex) \gets \mathcal{N}(\actoroutputsampled(\timeindex)) \) \Comment*[l]{Sample with variance}
	}
	{
		\( \actoroutputsampled(\timeindex) \gets \actoroutputmeans(\timeindex) \)
	}
	
	\( \precodingmatrix(\timeindex) \gets \text{OutputTransform}(\actoroutputsampled(\timeindex)) \)
	
	\( \sumrate(\timeindex) \gets \text{Evaluate}(\precodingmatrix(\timeindex)) \)
	\Comment*[l]{Acc. to \refeq{eq:sumrate}}
	
	Save data tuple\( (\statevec'(\timeindex), \actoroutputsampled(\timeindex), \sumrate(\timeindex) ) \)
	
	\( \timeindex \gets \timeindex + 1 \)%
\end{algorithm}

\begin{algorithm}[t]
	\caption{Learning Step}%
	\label{alg:learning}%
	\KwData{Experience Buffer}%
\KwResult{Parameter updates to \acs{acnn}, \acs{crnn}}%
\If{\( \timeindex \text{ mod } \stepsperupdate = 0 \)}
{
	Draw Batch~\( \batchset \) from Experience Buffer
	
	\Comment*[l]{Train Critics}
	
	\For{\( \idx \in \{ 1, 2 \} \)}
	{
		\For{\( \batchindex \in \{1, \ldots, \batchsize\} \)}
		{
			\(\hat{\sumrate}_{\batchindex} \gets \criticfunction_{\idx}(\statevec_{\batchindex}, \actoroutputsampled_{\batchindex}) \)
		}
		\( \loss'_{\criticfunction, \idx} \gets \) BatchLoss\((\hat{\mathbf{\sumrate}})\)
		\Comment*[l]{Acc. to \refeq{eq:losscriticfinal}}
		
		\( \paramscriticnum{\idx} \gets \)\acs{sgd}Step\((\loss'_{\criticfunction, \idx})\)
		\Comment*[l]{Acc. to \refeq{eq:sgdcritic}}
	}
	
	\Comment*[l]{Train Actor}
	
	\For{\( \batchindex \in \{1, \ldots, \batchsize\} \)}
	{
		\( \actoroutputsampled \gets \actorfunction(\statevec_{\batchindex}) \)
		\Comment*[l]{Get action according to current \acs{acnn}}
		
		\For{\( \idx \in \{ 1, 2 \} \)}
		{
			\( \hat{\sumrate}_{\idx} \gets \criticfunction_{\idx}(\statevec_{\batchindex}, \actoroutputsampled) \)
		}
	}
	\( \loss'_{\actorfunction} \gets \) BatchLoss\( (\hat{\mathbf{\sumrate}}_{1}, \hat{\mathbf{\sumrate}}_{2}) \)
	\Comment*[l]{Acc. to \refeq{eq:lossactorfinal}}
	
	\( \paramsactor \gets \)\acs{sgd}Step\((\loss'_{\actorfunction})\)
	\Comment*[l]{Acc. to \refeq{eq:sgdactor}}
}%
\end{algorithm}

\section{ANALYTICAL BASELINE PRECODERS}
\label{sec:baselineprecoders}

As noted by, e.g., \cite{perez2019signal}, existing precoder solutions for satellite downlink are either high-complexity or not sum rate optimal, and are not typically designed for robustness.
In this work, we will look at two analytical precoder baselines to benchmark the learned precoders' performance.
First, the \ac{mmse} precoder~\cite{chatzinotas2011energy} is the \ac{mimo} precoding workhorse, offering relatively low complexity, while not inherently accounting for inaccurate \ac{csit} estimations.
Additionally, for single-satellite scenarios we implement a more robust analytical precoder design that maximizes the \ac{slnr}~\cite{roper2023robust} as a proxy to the intractable \ac{sinr}~\refeq{eq:ergodicsumrate} given the first error model as defined above, thereby including robustness into its design.
Though the \ac{mmse} precoder in particular is frequently used, neither precoder directly maximizes the sum rate~\refeq{eq:optimizationobjective} in general.
Both precoders \emph{do} maximize the sum rate under certain conditions~\cite{roper2023robust}, primarily depending on accurate \ac{csit} and orthogonal channels, and significant spatial separation may be assumed to achieve approximately orthogonal channels.
However, upon violating these special conditions, both precoders' sum rate performance may degrade quickly, as we see in the evaluation.
Further, extending the robust \ac{slnr} precoder to both multi-satellite and multi-user is not trivial.
Complications such as these, namely the intractability of directly maximizing the sum rate analytically and inflexibility in system assumptions, are part of the motivating factors for exploring the use of \ac{ml} techniques.
In the following, we introduce mathematically the implementation of the \ac{mmse} and robust \ac{slnr} precoders for the given scenario.

First, the \ac{mmse} precoder finds its precoding~\( \precodingmatrixmmse \) such that
\begin{align}
	\precodingmatrixmmse &=
		\sqrt{
			\frac{\txpower}{\text{tr}\{\precodingmatrix'^{\text{H}}\precodingmatrix'\}}
		}
		\cdot \precodingmatrix'
	\\
	\text{with}\quad
	\precodingmatrix' &= \left[
	\csimatrixestim^{\text{H}}\csimatrixestim + \noisepower \numusers / \txpower \cdot \mathbf{I}_{\numsats\numantennaspersat}
	\right]^{-1}
	\csimatrixestim^{\text{H}}
	,
\end{align}
balancing channel distortion and noise power.
For a fair comparison, we also normalize each satellite's precoding slice to have equal maximum power \( \txpower_{\satelliteindex} \leq \txpower/\numsats \), same as the learned precoders.
The robust \ac{slnr} precoder uses the user~\( \userid \)'s channel power~\( \powerchannelslnr_{\userid} \) and the steering vector's autocorrelation~\( \steeringautocorrelation_{\userid} \) to determine its precoding slices~\( \precodingvectorslnr \) as
\begin{align}
	\precodingvectorslnr = 
		\frac{\txpower}{\numusers} \maxeigslnr,
\end{align}
where \( \maxeigslnr \) is the eigenvector that corresponds to the largest eigenvalue of
\begin{align}
	\left(
		\sum_{\userid'\in\userset, \userid'\neq\userid}
			\powerchannelslnr_{\userid'}^{2} \steeringautocorrelation_{\userid'}
			+ \noisepower\numusers / \txpower \cdot \mathbf{I}_{\numantennaspersat}
	\right)^{-1}
	\powerchannelslnr_{\userid}^{2} \steeringautocorrelation_{\userid}.
\end{align}
The robust \ac{slnr} precoder design assumes an error that affects the steering vector, such as the first error model~\refeq{eq:errormodel1}.
Recall that this error model causes the antennas of the satellites' \acp{ula} to have differently scaled errors depending on their positions.
Hence, using the steering autocorrelation~\( \steeringautocorrelation_{\userid} \) effectively distributes the transmit power to antennas less affected by error, thereby creating wider beams.
To calculate the autocorrelation~\( \steeringautocorrelation_{\userid} \), the error statistics, i.e., the error's distribution and its parametrization, must be known in advance.
The precoding slices~\( \precodingvectorslnr \) are then stacked as ~\( {\precodingmatrixslnr = [ \precodingvector_{\text{SLNR}, \num{1}}, \ldots, \precodingvector_{\text{SLNR}, \numusers} ]}, { \precodingmatrix_{\text{SLNR}}\in\numberscomplex^{\numsats\numantennaspersat\times\numusers} } \).

\subsection{Adaptations for Distributed Precoding}

We use the \ac{mmse} precoder as distributed precoding performance benchmark.
For the local case \refeq{eq:csiblind}, each satellite~\( \satelliteindex \) calculates their precoder slice~\( {\precodingmatrix_{\text{local}, \satelliteindex} \in \numberscomplex^{\numantennaspersat\times\numusers} }\) using only local information \( \csimatrixestim_{\satelliteindex} \) as
\begin{align}
	\precodingmatrix_{\text{local}, \satelliteindex} &= 
		\sqrt{
			\frac{\txpower/\numsats}{\|\precodingmatrix_{\text{local}, \satelliteindex}'\|^{2}}
		}
		\cdot \precodingmatrix_{\text{local}, \satelliteindex}' \\
	\text{with}\quad
		\precodingmatrix_{\text{local}, \satelliteindex}' &= \left[
		\csimatrixestim_{\satelliteindex}^{\text{H}}\csimatrixestim_{\satelliteindex} + \noisepower \numusers / \txpower \cdot \mathbf{I}_{\numsats\numantennaspersat}
		\right]^{-1}
		\csimatrixestim_{\satelliteindex}^{\text{H}}
		.
\end{align}
In case limited information is available at satellite~\(\satelliteindex\), i.e., \(\csimatrixestimdecone\)~\refeq{eq:csilimited1} or \( \csimatrixestimdectwo \)~\refeq{eq:csilimited2}, each satellite calculates its precoding slice~\( \precodingmatrix_{\text{L1}, \satelliteindex}, \precodingmatrix_{\text{L2}, \satelliteindex} \) as follows,
\begin{align}
	\precodingmatrix_{\text{L1}, \satelliteindex} &=
		\csimatrixestim_{\satelliteindex}
		\left[
			\csimatrixestimdecone\csimatrixestimdecone^{\text{H}}
			+
			\noisepower \numusers / \txpower \cdot \mathbf{I}_{\numusers}
		\right]^{-1}
	,\\
	\precodingmatrix_{\text{L2}, \satelliteindex} &=
		\csimatrixestim_{\satelliteindex}
		\left[
		\csimatrixestimdectwo\csimatrixestimdectwo^{\text{H}}
		+
		\noisepower \numusers / \txpower \cdot \mathbf{I}_{\numusers}
		\right]^{-1}
	.
\end{align}
The precoding slices~\(\precodingmatrix_{\text{local}, \satelliteindex}\) are stacked to form a full precoding matrix~\( {\precodingmatrix_{\text{local, MMSE}} = [\precodingmatrix_{\text{local}, 1}^{\text{T}}, \ldots, \precodingmatrix_{\text{local}, \numsats}^{\text{T}}]^{\text{T}}} \) and analogously for \( \precodingmatrix_{\text{L1, MMSE}}, \precodingmatrix_{\text{L2, MMSE}} \).
The full precoding matrix is evaluated according to \refsec{sec:setup}, the results of which we analyze alongside all other introduced precoders and scenarios in the upcoming section.


%


\section{EVALUATION}
\label{sec:experiments}
With both the data-driven precoder learning process and the benchmark precoders introduced, we now evaluate and compare them on a selection of scenarios.
Evaluation of learned precoders is done after a learning section is concluded, i.e., \( \stepsperupdate \rightarrow \infty \) and the parameters are not updated during evaluation.
Successful convergence of the learning algorithm is assumed once no improvement in mean performance is observed over an extended period. 
We investigate single satellite and multi-satellite scenarios, primarily in terms of sum rate under increasingly unreliable \ac{csit}.
We aim to present scenarios where the benchmark precoders perform strongly as well as scenarios where the benchmarks struggle to achieve high sum rates, as described in the preceding \refsec{sec:baselineprecoders}.
We show that the learning process adapts well to the diverse scenarios and level of availability of information.
In particular, we show that decreasing reliability of \ac{csit} during training leads to increasingly robust precodings, at the cost of peak performance.

\subsection{Implementation Details}
\label{sec:implementationdetails}

\begin{table}[!t]
	\renewcommand{\arraystretch}{1.3}%
	\caption{List of System Parameters}%
	\label{tab:sysparams}%
	\centering%
	\setlength{\tabcolsep}{3pt}%
	\rowcolors{2}{white}{uniblue1!10}%
	
\begin{tabular}{p{70pt}p{40pt}p{70pt}p{40pt}}
	\hline
	Noise Power \(\noisepower\) 				& \(6\text{e-}13\)\,W	 	&  Tx Power \(\txpower\) 						& \SI{100}{\watt}\\
	Satellite Altitude \( \distance_{0} \) 		& \SI{600}{\km} 			& Antennas \( \numantennaspersat \) 			& \( \{2, 16, 32\} \) \\
	Users \(\numusers\) 						& \( \{3, 6, 9\} \) 					& Satellites \( \numsats \) 					& \( \{1, 2\} \)\\
	Wavelength \( \wavelength \) 				& \( \text{c}/\SI{2}{\giga\hertz} \) 					& LS Fading Scale \( \largescalesigma \) 		& \( \num{0.1}\)\\
	User Gain \( \usergain \) 					& \(0\)\,dBi				& Satellite Gain \( \satgain \) 				& \(20\)\,dBi\\
	\acs{ula} spacing \( \distance_{\antennaid} \) & \(3/2\wavelength\) 					& 	 		&  \\
	Avg. User Dist \( \distance_{\userset} \) 		& \multicolumn{3}{l}{\( \{\SI{1}{\kilo\meter}, \SI{10}{\kilo\meter}, \SI{100}{\kilo\meter}\} \)}\\
	User Dist. Roam \( \wiggleusr \) 					& \multicolumn{3}{l}{\( \{ \SI{500}{\meter}, \SI{5}{\kilo\meter}, \SI{50}{\kilo\meter} \} \)}\\
	Avg. Sat. Dist \( \distance_{\satelliteset} \) 	& \multicolumn{3}{l}{\( \{\SI{10}{\kilo\meter}, \SI{100}{\kilo\meter}\} \)}\\
	Sat. Dist. Roam \( \wigglesat \) 					& \multicolumn{3}{l}{\(\SI{0}{\meter} \)}\\
	\hline
\end{tabular}%
\end{table}

\begin{table}[!t]
	\renewcommand{\arraystretch}{1.3}%
	\caption{List of \ac{rl} Parameters}%
	\label{tab:learningparams}%
	\centering%
	\setlength{\tabcolsep}{3pt}%
	\rowcolors{2}{white}{uniblue1!10}%
	\begin{tabular}{p{70pt}p{40pt}p{70pt}p{40pt}}
	\hline
	$\phantom{}^{1}$Entropy Scale \( \entropyscale \) 				& \num{1.0}					& $\phantom{}^{1}$Min. Samples \( \minsamples \) 					& \num{1000}\\
	$\phantom{}^{1}$Batch Size \( \batchsize \) 					& \num{1024} 				& $\phantom{}^{1}$Buffer Size \( \expbuffersize \) 					& \num{100000}\\
	$\phantom{}^{1}$L2 Actor \( \weightregulscaleactor \) 			& \num{0.01} 				& $\phantom{}^{1}$L2 Critic \( \weightregulscalecritic \) 			& \num{0.01}\\
	$\phantom{}^{1}$Training Interval \( \stepsperupdate \) 		& 10 						& $\phantom{}^{1}$Optimizer 										& Adam \cite{kingma2014adam} \\
	$\phantom{}^{2}$Actor \ac{lr} \( \learningrateactor \) 			& \num{4.2e-05} 					& $\phantom{}^{2}$Critic \ac{lr} \( \learningratecritic \) 			& \num{8.8e-06}\\
	$\phantom{}^{1}$Training Episodes		 						& \num{13000} 				& $\phantom{}^{1}$Steps per Episode 		 						& \num{1000}\\
	$\phantom{}^{1}$Actor Activation  		& \multicolumn{3}{l}{Penalized Tanh \cite{eger2019time}}\\
	$\phantom{}^{1}$Critic Activation  		& \multicolumn{3}{l}{Leaky ReLU \cite{eger2019time}}\\
	$\phantom{}^{1}$Actor Network Size  		& \multicolumn{3}{l}{\( [\num{2}\numusers\numsats\numantennaspersat, 512, 512, 512, 512, \num{4}\numusers\numsats\numantennaspersat] \)}\\
	$\phantom{}^{1}$Critic Network Size  		& \multicolumn{3}{l}{\( [\num{4}\numusers\numsats\numantennaspersat, 512, 512, 512, 512, 1] \)}\\
	\hline
\end{tabular}
\raggedright
\vspace{0.1ex}\\
$\phantom{}^{1}$Parameter is set using reasonable defaults.
$\phantom{}^{2}$Parameter is subsequently set by TPESampling search using the Optuna library~\cite{optuna_2019}.%
\end{table}

\reftab{tab:sysparams} and \reftab{tab:learningparams} list the selected system and learning parameters.
In multi-satellite scenarios, a reduced number of total antennas is sufficient since the wide inter-satellite distance \( \distance_{\satelliteset} \gg \distance_{\antennaid} \) allows narrow beams even with a low number of antennas.
We primarily evaluate the achieved sum rate~\refeq{eq:sumrate} averaged over \num{5000} Monte Carlo evaluations per evaluation point.
The rate that a precoder achieves for an evaluation varies not just due to the stochastic components of the error, but also due to the relative positioning of satellites and users.
In general, when user channels are more highly correlated, the achievable rate for a specific evaluation is lowered.
While different precoding strategies significantly affect the mean achieved sum rate, their variance around the mean is primarily determined by the realizations of positioning and error rather than by choice of algorithm.
Hence, we omit presenting the variance for readability.
The full code implementation and trained models are provided online~\cite{code}.
We also note that the selected learning parameters do not represent the optimum, rather, we set most training parameters to reasonable defaults and use the Optuna library~\cite{optuna_2019} to search for good solutions for the remaining parameters, as highlighted in \reftab{tab:learningparams}.
More thorough learning parameter optimization, while not within the scope of this paper, would yield slightly more effective precoding algorithms, though the presented results sufficiently show the potential of the approach.
The parameters will also have to be adapted to the specific deployed hardware.

During training, the learned precoding strategies are evaluated over batches of simulation steps~\( \timeindex \) due to the non-stationary nature of the satellite downlink model.
We denote these batches as \emph{episodes}.
Thus, the total number of simulation steps during one learning process is the number of episodes multiplied by the number of steps per episode.

Though we do not enumerate all learned precoders for brevity, one or multiple distinct strategies are learned for each scenario.
We indicate precoders that learned from perfect \ac{csit} by blue color and square markers, and precoders that learned from imperfect \ac{csit} likewise by red colors and diamond markers.
Further, where multiple strategies are learned for a scenario, we mention the operating point during training in the legend, i.e., “\( \text{SAC} \Delta\epsilon = 0.25 \)'' is trained with data generated at an error bound of \( {\erroraodbound=0.25} \).
Note that the axis scaling differs between plots of different scenarios to highlight the most relevant cross-section of data.

\subsection{Results}
\label{sec:results}

\subsubsection{Centralized Precoding}
We begin with a single satellite scenario with wide mean inter-user spacing of \( {\distance_{{\userset}}=\SI{100}{\kilo\meter}} \), $\numusers =3$~users and $\numantennaspersat=16$~satellite antennas.
This scenario is selected to enable strong performance in the baseline precoders, which approach the optimal sum rate for without error (\ac{mmse}) and including error (robust \ac{slnr}).
As the system model describes, each precoder is presented with a certain \ac{csit} estimate~\( \csimatrixestim \) and selects a precoding~\( \precodingmatrix \) based on it.
The precoding is evaluated in terms of sum rate \(\sumrate\) achieved when a precoding \( \precodingmatrix \) is applied to the true \ac{csit}~\( \csimatrix \).
We compare the \ac{mmse} and robust \ac{slnr} precoders to multiple learned precoding strategies \ac{sac}, and evaluate the precoders' performance over an increasing error bound~\( \erroraodbound \) according to the error model presented in \refsec{sec:setup}.

\begin{figure}
	\centering%
		\includegraphics{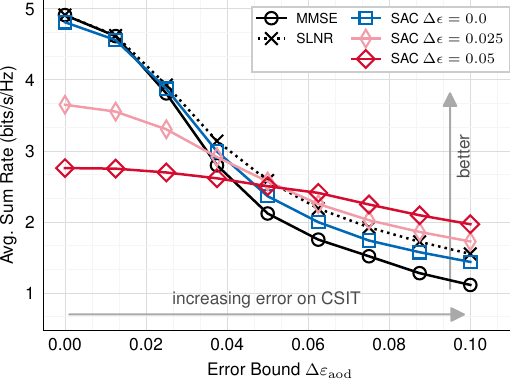}
	\caption{%
		Average precoding sum rate performance in single satellite downlink at mean inter-user distance \( \distance_{\userset} = \SI{100}{\kilo\meter}, \wiggleusr = \SI{50}{\kilo\meter} \) over increasingly unreliable \ac{csit}.
		Three learned algorithms \ac{sac} trained with perfect or increasingly unreliable information show elevated robustness over analytical baseline algorithms \ac{mmse}, \ac{slnr}.
	}
	\label{fig:1satstrongbaselines}
\end{figure}

\reffig{fig:1satstrongbaselines} shows the mean achieved sum rates.
We see strong performance from the baseline precoders.
Among the learned precoders, the precoder learned with perfect \ac{csit}, i.e., \(\erroraodbound=0\), only barely approaches the analytical precoders' performance.
As the \ac{csit} degrades, we see that this learned precoder maintains strong performance, showing increased robustness over the \ac{mmse} baseline precoder, though slightly falling behind the robust \ac{slnr} baseline precoder.
With the two robust learned algorithms \(\Delta\epsilon=0.025\) and \(\Delta\epsilon=0.05\), we also see that, by using increasingly unreliable training data, we can produce increasingly robust learned algorithms.
For the given error model, antennas are afflicted with increasingly severe error realizations the further away they are from the center of the antenna array.
The robust \ac{slnr} precoder, as described in the preceding section, effectively uses this information to allocate more power to antennas closer to the center.
Looking at the robust learned precoder \ac{nn}'s input weights, we infer that it has also learned to focus the center antennas, allocating on average about \SI{23}{\percent} more weight to phase information from the inner third of antennas compared to the outer thirds.

\begin{figure}
	\centering%
		\includegraphics{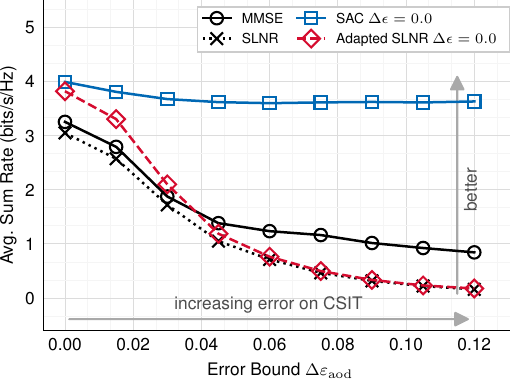}
	\caption{%
		Average precoding sum rate performance in single satellite downlink at mean inter-user distance \( \distance_{\userset} = \SI{10}{\kilo\meter}, \wiggleusr = \SI{5}{\kilo\meter} \) over increasingly unreliable \ac{csit}.
		The learned \ac{sac} precoder remains robust, while analytical baselines degrade with higher user correlation.
		As a hybrid approach, the adapted \ac{slnr} closes the \ac{slnr}'s performance gap at the training point of \( \Delta\epsilon=0.0 \).
	}
	\label{fig:1satweakbaselines}
\end{figure}

Moving on to scenarios that less favor the baseline algorithms, \reffig{fig:1satweakbaselines} shows performance in a single satellite scenario with a relatively closer mean inter-user distance of \( {\distance_{{\userset}}=\SI{10}{\kilo\meter}} \).
In the single satellite scenario, the baseline algorithms struggle with these closer inter-user distances.
The \ac{slnr} precoder in particular, being used outside its design space, no longer provides robustness benefits even over the \ac{mmse} precoder.
\ac{sac}, even with perfect \ac{csit} during training, learns a strategy that outperforms both baseline algorithms as well as showing strong robustness.

\begin{figure}
	\centering%
		\includegraphics{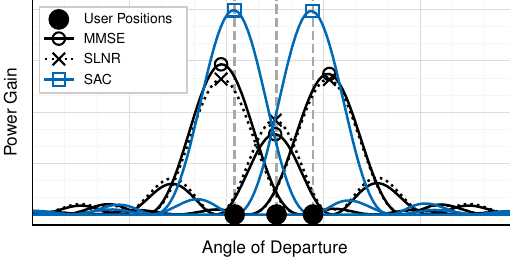}
	\caption{%
		Sample beam patterns for a constellation with a single satellite downlink and relatively low mean inter-user distance \( \distance_{\userset} = \SI{10}{\kilo\meter} \).
		Achieved sum rates: \ac{mmse}: \num{3.7} bps/Hz, \ac{slnr}: \num{3.6} bps/Hz, \ac{sac}: \num{4.5} bps/Hz.
		At this beam-width, \ac{mmse} and \ac{slnr} precoders struggle to resolve the three users.
		The learned algorithm leverages power allocation to serve primarily the outer two users, which is more beneficial to maximizing sum rate in this case.
	}
	\label{fig:beampattern_1sat}
\end{figure}

\reffig{fig:beampattern_1sat} shows a sample beam pattern for each precoder, providing intuition on why the analytical baselines struggle to achieve high sum rates with close inter-user distances.
With the relatively wide beams of the single satellite setup, resolving all three users satisfactorily is only possible at the cost of high interference to the center user.
Instead, the \ac{ml} algorithms learn to leverage power allocation more aggressively, serving primarily the outer users, which turns out to be more conducive to achieving high sum rates in this scenario.
If fairness is a consideration, the learning algorithm could be adapted to maximize a mixed performance objective, as other work, e.g., \cite{alsenwi2023robust}, have shown.

\begin{figure}
	\centering%
	\includegraphics{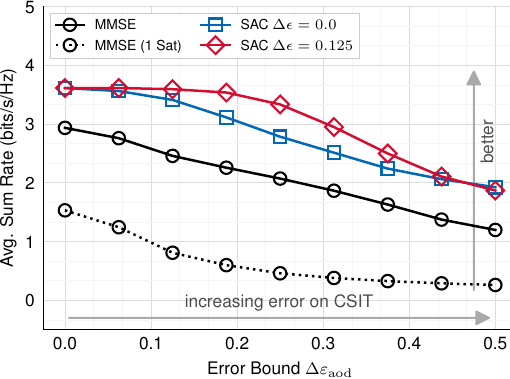}
	\caption{%
		Average precoding sum rate performance in two satellite downlink at inter satellite distance of \( \SI{100}{\kilo\meter} \) and mean inter-user distance \( \distance_{\userset} = \SI{1}{\kilo\meter}, \wiggleusr = \SI{500}{\meter} \) over increasingly unreliable \ac{csit}.
		Cooperative two-satellite precoding achieves significantly higher sum rates than single satellite \ac{mmse}, with the learned \ac{sac} algorithms showing superior robustness and overall performance.
	}
	\label{fig:2sat}
\end{figure}

In the two satellite scenario shown in \reffig{fig:2sat}, all precoders including the analytical baselines achieve a significant sum rate benefit using two cooperative satellites over using a single satellite.
However, the learned algorithms still outperform the \ac{mmse} baseline, and the robust learned precoder maintains strong sum rate performance even at high levels of unreliability in \ac{csit}.

\begin{figure}
	\centering%
	\includegraphics{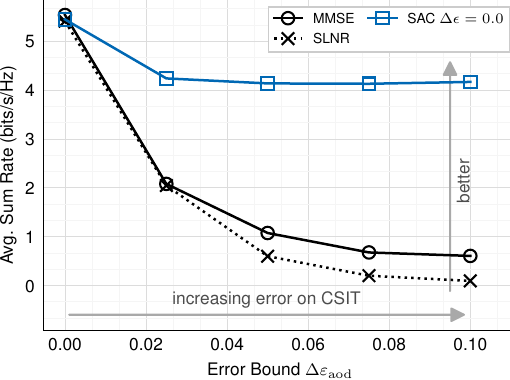}
	\caption{%
		Average precoding sum rate performance in single satellite downlink for more complex scenario with \( \numusers=6 \) users, \( \numantennaspersat=32 \) antennas at mean inter-user distance \( \distance_{\userset} = \SI{10}{\kilo\meter}, \wiggleusr = \SI{5}{\kilo\meter} \) over increasingly unreliable \ac{csit}.
		The learned algorithms deal well with the increased complexity of the scenario.
	}
	\label{fig:1sat6usr}
\end{figure}

\begin{figure}
	\centering%
	\includegraphics{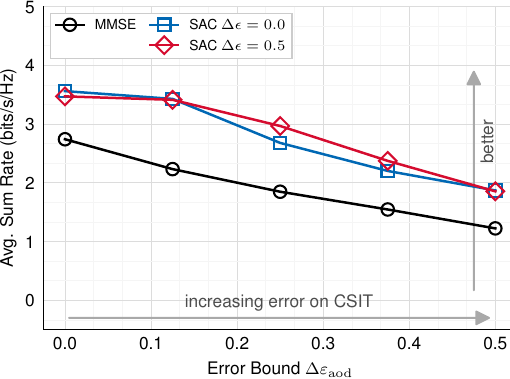}
	\caption{%
		Average precoding sum rate performance in multi satellite downlink for more complex scenario with \( \numusers=9 \) users at mean inter-user distance \( \distance_{\userset} = \SI{1}{\kilo\meter}, \wiggleusr = \SI{500}{\meter} \) over increasingly unreliable \ac{csit}.
		The learned algorithms deal well with the increased complexity of the scenario.
	}
	\label{fig:2sat9usr}
\end{figure}

To confirm that the learning method scales to more complex scenarios, we double the number of antennas and number of users for the single satellite scenario, and triple the number of users for the multi-satellite scenario.
The irregular antenna array spanned by multiple satellites allows fairly tight beams with a relatively low number of antennas.
The results, shown in \reffig{fig:1sat6usr} for single satellite and \reffig{fig:2sat9usr} for multi-satellite, show that the learning method maintains performance for these more complex scenarios.

\subsubsection{Distributed Precoding}
\begin{figure}
	\centering%
	\includegraphics{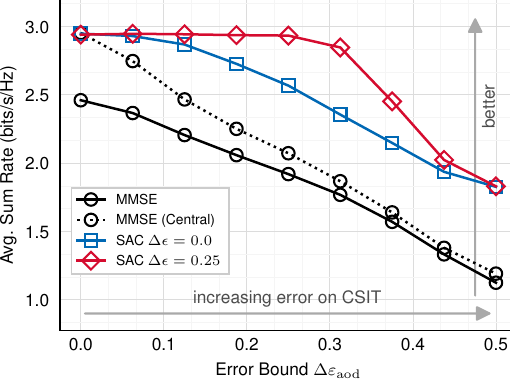}
	\caption{%
		Average precoding performance for multi-satellite cooperative downlink with only local information available at each satellite.
		Inter satellite distance of \( \SI{100}{\kilo\meter} \) and mean inter-user distance \( \distance_{\userset} = \SI{1}{\kilo\meter}, \wiggleusr = \SI{500}{\meter} \) over increasingly unreliable \ac{csit}.
		Although all precoders degrade under local information, the learned algorithms retain their performance and robustness advantage.
	}
	\label{fig:decentralizedblind}
\end{figure}

Here, the satellites~\( \satelliteindex \) are provided with either fully local \ac{csit} estimates~\( \csimatrixestim_{\satelliteindex} \)~\refeq{eq:csiblind} or limited information of the other satellites channels, \( \csimatrixestimdecone \)~\refeq{eq:csilimited1} or \( \csimatrixestimdectwo \)~\refeq{eq:csilimited2}.
From this, each satellite~\( \satelliteindex \) locally builds a precoding slice~\( \precodingmatrix_{\satelliteindex} \).
The precoding slices are stacked and the full precoding matrix~\( \precodingmatrix \) is evaluated on the true \ac{csit} in terms of achieved sum rate~\( \sumrate \).
\reffig{fig:decentralizedblind} shows the fully local scenario.
While all precoders suffer from a performance dip dealing with only local information compared to the centralized precoders, the learned precoders retain their positive performance gap over the \ac{mmse} baseline.
In particular, the local learned precoders manage to outperform even the centralized \ac{mmse}.
Significant robustness gains are achieved by adding unreliable \ac{csit} samples to the learning process.

\begin{figure}
	\centering%
	\includegraphics{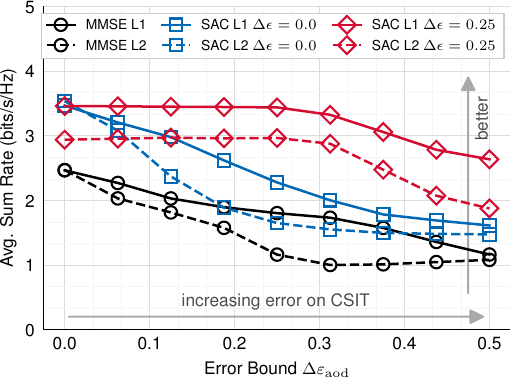}
	\caption{%
		Average precoding performance for multi-satellite cooperative downlink with only limited outside information available at each satellite.
		Inter satellite distance of \( \SI{100}{\kilo\meter} \) and mean inter-user distance \( \distance_{\userset} = \SI{1}{\kilo\meter}, \wiggleusr = \SI{500}{\meter} \) over increasingly unreliable \ac{csit}.
		L1 (straight lines) and L2 (dashed lines) correspond to the "Limited 1" and "Limited 2" local information models.
	}
	\label{fig:decentralizedlimited}
\end{figure}

In \reffig{fig:decentralizedlimited} we evaluate performance for local precoding with limited knowledge of other satellites' \ac{csit}.
As described above, we simulate outdated \ac{csit} by giving each satellite perfect own \ac{csit} and erroneous \ac{csit} of the other satellite (L1) or erroneous own \ac{csit} and scaled erroneous \ac{csit} of other satellites (L2).
Mirroring centralized precoding, significant sum rate and robustness gains are achieved over the baselines.
Though each satellite precodes independently, the performance of centralized precoding is matched at perfect \ac{csit}, i.e., \( \erroraodbound=0.0 \), compared to \reffig{fig:2sat}.
At L1 information, the robust learned precoders manage to even outperform the centralized precoder, which is attributed to having error-free own channel estimates available.
We can trace the use of this information in the \ac{nn} learned parameters, where the own phase information receives around \SI{25}{\percent} higher weight compared to the other satellite's phase information.
At L2 information, this advantage disappears and performance drops for all methods. However, our \ac{sac}-based algorithms, particularly the one trained with imperfect \ac{csit}, maintain a substantial performance advantage over the \ac{mmse} baseline.
Overall, the learned algorithms adapt well to the given level of information even in multi-agent learning scenarios.
\subsubsection{Mixed Error Models}
\begin{figure}
	\centering%
	\includegraphics{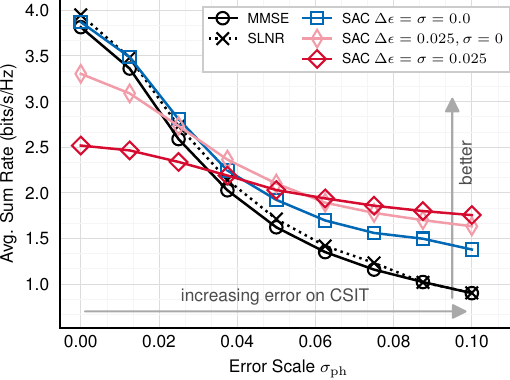}
	\caption{%
		Average precoding sum rate performance in single satellite downlink at mean inter-user distance \( \distance_{\userset} = \SI{100}{\kilo\meter}, \wiggleusr = \SI{50}{\kilo\meter} \) over increasingly unreliable \ac{csit}.
		Mixed error model, where the error scale \( \errorphasevariance \) is sweeped while the other error source is active at a constant bound \( \erroraodbound=\num{0.025} \).
		The learned models show increased robustness compared to the baselines \ac{mmse}, \ac{slnr}, which provide little to no robustness on this mixed error model that lies outside their design space.
	}
	\label{fig:1satothererror}
\end{figure}

Under real-life conditions, error models may not grasp the full picture.
As we have seen earlier, analytically derived algorithms' performance can degrade quickly when used outside their design space, and this holds true for the error modeling as well.
We demonstrate this by applying a mixed error model as described in \refsec{sec:setup}, where we apply both positioning and phase error at the same time.
We hold the positioning error bound constant at \(\erroraodbound=0.25\) and evaluate precoders at increasing phase error scale \( \errorphasevariance \), presented in \reffig{fig:1satothererror}.
As defined earlier, the \ac{mmse} precoder has no inherent robustness mechanism, while the \ac{slnr} precoder, as described in this work, is designed for positioning errors \( \erroraod \), but not phase errors \( \errorphase \).
As such, neither analytical baseline provides much robustness as the influence of the phase error \( \errorphase \) increases.
Much like before, learned algorithms can be designed with increasing levels of robustness by presenting them with increasingly unreliable training data.
\subsubsection{Adapting Model-Based Precoders}
\begin{figure}
	\centering%
	\includegraphics{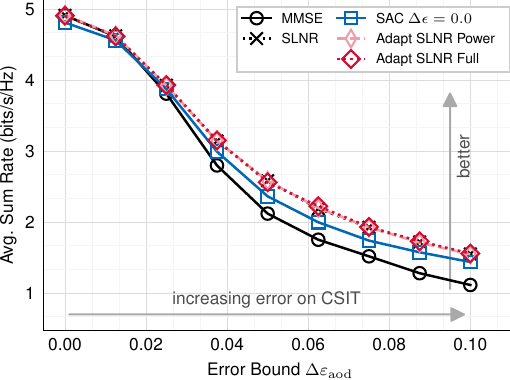}
	\caption{%
		Average precoding sum rate performance in single satellite downlink at mean inter-user distance \( \distance_{\userset} = \SI{100}{\kilo\meter}, \wiggleusr = \SI{50}{\kilo\meter} \) over increasingly unreliable \ac{csit}.
		Both learned precoders that adapt the \ac{slnr} precoder are flush in performance with it, equalizing the performance gap observed from pure learning (blue square).
	}
	\label{fig:1satadaptnoerror}
\end{figure}

As we see, though the learned precoding strategies outperform the model based precoder baselines in aggregate, the analytical precoders work well at their design point as seen in, e.g., \reffig{fig:1satstrongbaselines}.
For the last evaluation, we want to take a short peek at the possibility of using existing precoders as a starting point for the learned precoding strategies.
To highlight this possibility, we switch the learned precoder to not directly output a precoding, but rather a matrix that scales 1)~the robust \ac{slnr} precoder's power allocation; or 2)~the entries of the robust \ac{slnr} precoding.
We evaluate this composite precoder in \reffig{fig:1satadaptnoerror} and \reffig{fig:1satweakbaselines} for the scenario where the robust \ac{slnr} precoder performs well and badly, respectively.
We see that in the scenario where the robust \ac{slnr} precoder already performed well, using it as a starting point allows the learning approach to better match the analytical baseline.
At the same time, in \reffig{fig:1satweakbaselines} the hybrid learning approach significantly closes the performance penalty that the analytical precoder suffers at the training point of \( \erroraodbound=0.0 \).
Though it does not generalize well outside that training point, it has extended the scope of the analytical precoder outside of its design space.
Thus, hybrid approaches may combine the benefits of analytical and data-driven design.

\begin{figure}
	\centering%
	\includegraphics{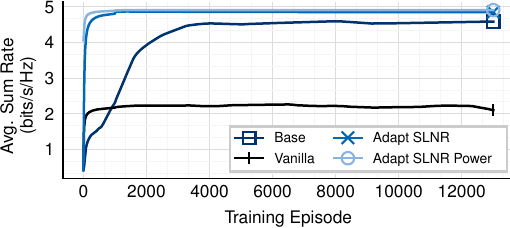}
	\caption{%
		Mean sum rate per episode achieved during training, averaged over a sliding window of length \num{1000}.
		Adapting the analytical \ac{slnr} benchmark by \ac{ml} methods leads to significantly faster and more stable training.
		Learning with a vanilla \ac{sac} algorithm without the adaptations presented in this paper converges to a local minimum solution significantly worse than other learned precodings.
	}
	\label{fig:training}
\end{figure}

\reffig{fig:training} displays a further big advantage of such combined approaches: convergence during training is significantly sped up by providing the \ac{sac} learning with a strong starting point.
While this particular implementation is not especially suited for real applications due to the need to evaluate both the robust \ac{slnr} precoder as well as the adapting \ac{nn} during inference, it highlights the potential of combining model-based precoders with learning methods for, e.g., continually learning systems.
\subsubsection{Ablation Study}
\label{sec:ablationstudy}
We train a vanilla \ac{sac} algorithm without the additions described in \refsec{sec:OURAPPROACH}, i.e., no low frequency network updates, no input standardization and inter layer normalization, and no weight penalty, on a scenario with \num{1} satellite and \SI{100}{\kilo\meter} mean inter user distance~\( \distance_{{\userset}} \).
As shown in \reffig{fig:training}, the vanilla \ac{sac} trained precoder reaches a low performance local minimum.
Though we cannot rule out that the vanilla \ac{sac} could perform adequately given the right set of hyperparameters, we have not found such a set, and it is our experience that the additions made convergence to a high performance minimum far less sensitive to hyperparameter selection.

\subsection{Computational Complexity}
Estimating the computational complexity for analytical algorithms in advance is fairly straight forward, i.e., a \ac{mmse} precoder is dominated by the matrix inversion, which is known to scale at about \( O((\numantennaspersat\numsats)^{3}) \), depending on the implementation.
With \ac{nn}, the computational complexity depends on the \ac{nn} model size, i.e., layers and nodes per layer.
Selecting the minimum model size that has the capacity to approximate the desired algorithm with sufficiently high fidelity is non-trivial and an unsolved problem in research.
Practical approaches are almost exclusively heuristic, e.g., \cite{cai2018proxylessnas} or \cite{kaveh2023application}.
Thus, the exact scaling with problem complexity is not evaluated in this work.
However, computational complexity is critical to judging the practical viability of a real time algorithm, so we give an indication on the expected complexity by comparing run times for the learned precoders that we use in this work, evaluated on a generic CPU (Intel i5-1235U).
With the inference as depicted in the flowchart in \reffig{fig:sacflowchart}, the only steps that the learned precoder must perform in real time on constrained satellite hardware are the evaluation of the \ac{acnn} and related pre- and postprocessing.
\reftab{tab:runtime} shows the median run time of \num{100000} evaluations of a learned precoder compared to the two analytical baselines for a selection of scenarios with increasing complexity.
The learned precoder is overall comparable to the \ac{mmse} precoder and much faster than the robust \ac{slnr} precoder, though the \ac{mmse} precoder shows an advantage on evaluations with less parameters.
In more general settings, we expect the computational complexity of the learned precoder to scale favorably for the following reasons:
1)~In this paper, we use a single \ac{acnn} model size that works for all evaluations.
Since some of the evaluations are strictly more complex than others, e.g., when the number of users and antennas increases, we can assume that the model is overdimensioned for at least some of the evaluations.
2)~Evaluation of fully connected \ac{nn} uses simple, highly parallelizable operations, such that availability of dedicated hardware on the satellite, i.e., GPU or NPU, could have a positive effect.
3)~As the learned algorithms are only approximately optimal, if computational resources are constrained, it is common practice to start with an overdimensioned \ac{nn} and compress the resulting learned algorithm by methods such as quantization or pruning.
Though we do not investigate compression in depth in this work, we note faster run times with no loss in performance when applying generic dynamic range quantization via \cite{tflite}.
In terms of storage complexity, the \ac{acnn} model used in all simulations uses at most \SI{5.5}{\mebi\byte} of memory, or \SI{1.4}{\mebi\byte} when compressed.

\begin{table}[!t]
	\renewcommand{\arraystretch}{1.3}%
	\caption{Median Precoding Run Time on generic CPU}%
	\label{tab:runtime}%
	\centering%
	\includegraphics{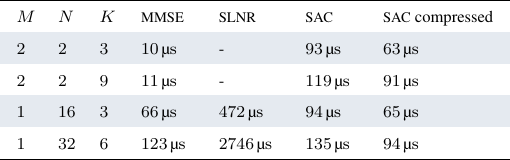}
	\setlength{\tabcolsep}{8pt}%
	\rowcolors{2}{white}{uniblue1!10}%
\end{table}

\subsection{Limitations}
\label{sec:limitations}
While a perfect learning process would result in the beamforming algorithm that is optimal at its training point, convergence and steerability remain an ongoing issue in \ac{dl} research~\cite{hendrycks2021unsolved}.
We see in, e.g., \reffig{fig:1satstrongbaselines} that neither learned algorithm achieves perfect performance at their training point, though generalization proceeds generally as expected, i.e., the learned algorithm trained under the most unreliable \ac{csit} exhibits the strongest robustness as error increases further.
While the learned algorithms perform strongly overall, performance gaps can be further lessened through more careful selection of training parameters. Minor artifacts such as these are to be expected given current approximate, stochastic, and regularized \ac{dl} methods.
In situations where a straightforward and highly effective analytical algorithm is available, it is more appropriate to select traditional methods rather than learning-based approaches.
For all other scenarios, selecting approximate learned precoding strategies offers a significant degree of flexibility at high performance.



%


\section{CONCLUSION}
\label{sec:conclusions}

\glsresetall  

In this work, we show that data-driven learning can be applied to satellite communications as an extremely flexible method of deriving approximately optimal precoding algorithms.
We use \ac{rl}, a process of automated scientific trial-and-error, to iteratively improve upon a precoding strategy until is approximately maximizes a given performance objective.
Primarily, this is useful as a one size fits all approach to deriving algorithms regardless of the existence of an accurate system model and mathematically tractable optimization problem.
Further, the resulting algorithms all share the same \ac{nn} structural setup, making a change of algorithm as simple as swapping a set of parameters.
We discuss what adjustments need to be made to off the shelf \ac{rl} methods for the satellite downlink and show on a variety of scenarios, including single and cooperative multi-satellite with local or global information, that \ac{rl} can be applied to learn precoding algorithms that achieve high sum rates.
Compared to analytical baselines, the learned algorithms match performance where the baselines are near optimal and significantly outperform them otherwise.
In particular, we show that the approximate nature of current \ac{dl} methods lends itself well to optimizing for robustness to uncertain information, which is of notable use for real-world scenarios.
With power efficient \ac{nn} hardware implementations (e.g., \cite{intel2024npu}) and dedicated \ac{nn} hardware onboard satellites (e.g., \cite{ghiglione2022opportunities}) coming into focus, learned precoding algorithms may present an attractive option for future \ac{ntn}.


%

\bibliographystyle{IEEEtran}%
{%
	\makeatletter  
	\clubpenalty=10000  
	\@clubpenalty=\clubpenalty
	\widowpenalty=10000
	\makeatother
	\bibliography{%
		ref/IEEEabrv,
		ref/references
	}%
}%
\newpage

\end{document}